\documentclass[10pt,preprint]{aastex}

\newcommand{\ha}{\,H$\alpha$}     

\newcommand{\arcs}{$^{\prime\prime}$}
\newcommand{\arcm}{$^{\prime}$}

\newcommand{\galex}{{\em GALEX}}

\usepackage{lscape}

\shorttitle{Ultraviolet Luminous Galaxies}
\shortauthors{Hoopes et al.}

\begin{document}

\title{The Diverse Properties of the Most Ultraviolet Luminous Galaxies Discovered by the {\em Galaxy Evolution Explorer}}

\author{
Charles G. Hoopes\altaffilmark{1}, 
Timothy M. Heckman\altaffilmark{1}, 
Samir Salim\altaffilmark{2},
Mark Seibert\altaffilmark{3}, 
Christy A. Tremonti\altaffilmark{4}, 
David Schiminovich\altaffilmark{5}, 
R. Michael Rich\altaffilmark{2}, 
D. Christopher Martin\altaffilmark{3}, 
Stephane Charlot \altaffilmark{6,7}, 
Guinevere Kauffmann \altaffilmark{6},
Karl Forster\altaffilmark{3},
Peter G. Friedman\altaffilmark{3},
Patrick Morrissey\altaffilmark{3},
Susan G. Neff\altaffilmark{8},
Todd Small\altaffilmark{3},
Ted K. Wyder\altaffilmark{3},
Luciana Bianchi\altaffilmark{1},
Jose Donas\altaffilmark{9},
Young-Wook Lee\altaffilmark{10},
Barry F. Madore\altaffilmark{11},
Bruno Milliard\altaffilmark{9},
Alex S. Szalay\altaffilmark{1},
Barry Y. Welsh\altaffilmark{12},
Sukyoung K. Yi \altaffilmark{10}}

\altaffiltext{1}{Department of Physics and Astronomy, The Johns Hopkins
University, Homewood Campus, Baltimore, MD 21218}

\altaffiltext{2}{Department of Physics and Astronomy, University of
California, Los Angeles, CA 90095}

\altaffiltext{3}{California Institute of Technology, MC 405-47, 1200 East
California Boulevard, Pasadena, CA 91125}

\altaffiltext{4}{Steward Observatory, University of Arizona, 933 North Cherry Avenue, Tucson, AZ 85721}

\altaffiltext{5}{Department of Astronomy, Columbia University, New York, NY 10027}

\altaffiltext{6}{Max-Planck-Institut fur Astrophysik, Karl-Schwarzschild-Str. 1, D-85748 Garching bei Munchen, Germany}

\altaffiltext{7}{Institut d'Astrophysique de Paris, UMR 7095, 98 bis Boulevard Arago, F-75014 Paris, France}

\altaffiltext{8}{Laboratory for Astronomy and Solar Physics, NASA Goddard
Space Flight Center, Greenbelt, MD 20771}

\altaffiltext{9}{Laboratoire d'Astrophysique de Marseille, BP 8, Traverse
du Siphon, 13376 Marseille Cedex 12, France}

\altaffiltext{10}{Center for Space Astrophysics, Yonsei University, Seoul
120-749, Korea}

\altaffiltext{11}{Observatories of the Carnegie Institution of Washington,
813 Santa Barbara St., Pasadena, CA 91101}

\altaffiltext{12}{Space Sciences Laboratory, University of California at
Berkeley, 601 Campbell Hall, Berkeley, CA 94720}

\begin{abstract}

We report on the properties of a sample of ultraviolet luminous
galaxies (UVLGs) selected by matching the {\it Galaxy Evolution
Explorer (GALEX)} All-sky Imaging and Medium Imaging Surveys with the
Sloan Digital Sky Survey Third Data Release. The overlap between these
two surveys is roughly 450 square degrees. Out of 25362 galaxies (with
SDSS spectroscopy) between $0.0<z<0.3$ detected by \galex, there are
215 galaxies with $L>2\times10^{10}$~L$_{\odot}$ at 1530~\AA\
(observed wavelength). The properties of this population are well
correlated with ultraviolet surface brightness. We find that the
galaxies with low UV surface brightness are primarily large spiral
systems with a mixture of old and young stellar populations, while the
high surface brightness galaxies consist primarily of compact
starburst systems, with an approximate boundary at a surface
brightness of $I_{1530}=10^8$~L$_{\odot}$~kpc$^{-2}$. The large
galaxies appear to be the high-luminosity tail of the galaxy star
formation function, and owe their large luminosity to their large
surface area.  In terms of the behavior of surface brightness with
luminosity, size with luminosity, the mass-metallicity relation, and
other parameters, the compact UVLGs clearly depart from the trends
established by the full sample of galaxies. The subset of compact
UVLGs with the highest surface brightness
($I_{1530}>10^9$~L$_{\odot}$~kpc$^{-2}$; ``supercompact UVLGs'') have
characteristics that are remarkably similar to Lyman Break Galaxies at
higher redshift. They are much more luminous (and thus have much
higher star formation rates) than typical local ultraviolet-bright
starburst galaxies and blue compact dwarf galaxies. They have
metallicities that are systematically lower than normal galaxies of
the same stellar mass, indicating that they are less chemically
evolved. In all these respects, they are the best local analogs for
Lyman Break Galaxies.

\end{abstract}


\section{Introduction}	

Over the past decade enormous progress has been made toward mapping
the cosmological history of star formation in the universe ({\it
e.g.,} Madau et al. 1996; Giavalisco et al. 2004). This has mainly
been accomplished using large samples of high-redshift galaxies
selected by their rest-frame ultraviolet (UV) colors ({\it e.g.,}
Steidel et al 1996, 2003; Dickinson et al. 2004). These surveys
indicate that the global star formation rate of the universe has been
in decline since $z\sim1-2$, and was generally constant at higher
redshift out to at least $z=6$ \citep{g04}.

This picture still contains some uncertainty resulting from several
factors. The star formation rate density at low redshift ($z=0-1$) has
been determined through different techniques (${\it e.g.,}$
\ha\ luminosity) than those used for higher redshift galaxies (${\it
e.g.,}$ rest-frame ultraviolet luminosity). These techniques are affected
differently by extinction and radiative transfer effects, and they
fundamentally probe star formation over different time scales.

One way around this problem is to obtain rest-frame ultraviolet (UV)
measurements for a large sample of galaxies at low redshift, enabling
the measurement of star formation rates using the same techniques that
are used at higher redshift. This requires a UV telescope in space
with a large field of view, something that has not been available
until the {\it Galaxy Evolution Explorer (GALEX)} mission
\citep{cm05}. \galex\ is obtaining UV fluxes for more than $\sim10^7$
galaxies in the redshift range of $0<z<2$. Initial results on the UV
luminosity density show strong evolution from $z=2$ to $z=0$, with the
strongest evolution occurring in the most UV luminous galaxies
\citep{ds05,arn05}. The fraction of galaxies with $L_{1530}>0.2$L$_{*,
z=3}$ fell by a factor of 30 from $z=1$ to $z=0$ (using L$_{*,
z=3}=6\times10^{10}$ L$_{\odot}$; Steidel et al. 1999).

These UV luminous galaxies at high redshift are more commonly called
Lyman Break Galaxies (LBGs; Steidel et al. 1999). These high-redshift
galaxies are so named because they are identified by the effects of
the Lyman break on their broadband colors \citep{sh93}. LBGs are UV
bright galaxies undergoing intense star formation with low to moderate
stellar masses (log $M_*=9.5$ to $11.0$ M$_{\odot}$), and are
candidates for the precursors of present-day elliptical galaxies (see
{\it e.g,} Giavalisco 2002). LBGs are common at $z>2$, and they are
clearly important as the sites of a significant fraction of all the
star formation in the universe. Since strong evolution has made
objects like LBGs extremely rare in the local universe, all of the
information on this important galaxy population has come from very
distant samples, which are inherently difficult to study. Thus, there
has been little detailed information available on the processes
driving the evolution of star formation in the population of LBGs.

Using local UV-bright starbursts as local analogs to LBGs has
contributed significantly toward understanding these objects
\citep{h98,m99}. However, local starbursts differ from LBGs in
important ways. Local starbursts are usually dwarf galaxies or small
(sub-kpc) regions in the nuclei of larger galaxies, while LBGs have
typical sizes of a few kpc \citep{ferg04}. Luminous local starbursts
are usually very dusty systems in which only a small fraction of the
UV light escapes, while LBGs with similar bolometric luminosities
(star formation rates) typically contain modest amounts of dust
(e.g. Reddy et al. 2006; Erb et al. 2006b).  Given these differences,
it is not clear that the conditions in local starbursts or the
triggers of star formation are identical to those in LBGs, and so
there is a need for better LBG analogs in the local universe.

Since LBGs are found in part by their large UV luminosity, LBG analogs
in the local universe should also be UV luminous. The large area UV
sky surveys being carried out by \galex\ provide an ideal data set for
finding rare UV luminous galaxies in the local universe. Heckman et
al. (2005; hereafter Paper I) described the properties of the most UV
luminous galaxies (UVLGs) in the local universe based on
cross-matching the initial \galex\ surveys with the Sloan Digital Sky
Survey (SDSS) first data release (DR1; Abazajian et al. 2003). The
UVLGs were composed of two basic types of galaxies: Large UVLGs, which
are characterized by lower UV surface brightness and high mass, and
compact UVLGs, which have higher UV surface brightness and lower
mass. Many of the compact UVLGs have properties very similar to LBGs.

Although this sample was very illuminating, several questions
remain. The extent of the similarity between the compact UVLGs and
LBGs is a crucial question. More generally, it is not known whether
these galaxies are truly a distinct population of objects in an
earlier phase of evolution, {\it i.e.,} remnants of the epoch of
galaxy formation, or whether they are simply the high end of the UV
luminosity function. Many of these questions could be better addressed
if there were more such galaxies available for study, so we present an
analysis of a larger sample of UVLGs, based on more recent \galex\ and
SDSS data.

\section{Data}	

\subsection{Ultraviolet Data}	

Since its launch in April, 2003, \galex\ has been conducting several
surveys of the UV sky. In this paper we make use of the \galex\
All-sky Imaging Survey (AIS) and Medium-Deep Imaging Survey (MIS). The
data were taken from the first public release of \galex\ data (GR1)
available at the Multimission Archive at Space Telescope
(MAST). Details on the \galex\ mission and surveys are given in
\citet{cm05}.

The \galex\ data include far-ultraviolet (FUV; $\lambda_{eff}=1528$~\AA,
$\Delta\lambda=268$~\AA) and near-ultraviolet (NUV;
$\lambda_{eff}=2271$~\AA, $\Delta\lambda=732$~\AA) images with a
circular field of view with radius $\sim38$\arcm. The spatial
resolution is $\sim5$\arcs. Details of the \galex\ satellite and data
characteristics can be found in \citet{pm05}. 

The data were processed through the \galex\ reduction pipeline at the
California Institute of Technology. The pipeline reduces the data and
automatically detects, measures, and produces catalogs of FUV and NUV
fluxes for sources in the \galex\ images.

\subsection{Optical data}	

The \galex\ catalogs were then matched to the SDSS Third Data Release
(DR3; Abazajian et al. 2005) spectroscopic sample. The area of the
overlap region between GR1 and DR3 is about 450 square degrees
\citep{b06}. The SDSS catalog provides (among many other available
parameters) $ugriz$ magnitudes, spectroscopic redshifts, concentration
parameters, observed half-light radii and model-fit exponential scale
lengths. To be included in our final matched catalog, we required that
each source have a spectroscopic redshift in the range $0<z<0.3$, and
that the SDSS source be spectroscopically classified as a galaxy,
excluding objects classified by the SDSS pipeline as QSOs or Type I
(broad line) AGN. The resulting GR1/DR3 sample contains 25362
galaxies. Of these, 18463 have 3$\sigma$ FUV detections. The remaining
galaxies were detected in the NUV images only.

With the distances estimated from the SDSS redshift, the FUV (and NUV)
luminosity for each galaxy is known. Following Paper I, galaxies with
$L_{1530}>2\times10^{10}$~L$_{\odot}$ qualify as UV luminous
galaxies\footnote{Throughout this paper we use $H_0=70$~km~s$^{-1}$,
$\Omega_m=0.3$, and $\Omega_{\Lambda}=0.7$.}, where $L_{1530}$ is the
luminosity at the observed wavelength of 1530~\AA. This luminosity is
$\sim5L_*$ for $z=0$ \citep{w05} and $\sim0.3L_*$ for LBGs at $z=3$
\citep{s99}. There are 235 galaxies in the GR1/DR3 sample that meet
this criterion. We then inspected the SDSS spectra of these galaxies
to eliminate broad line (Type I) AGN that were missed by the SDSS
pipeline, as well as objects with BL Lac-type spectra (UV bright but
with weak or non-existent emission lines). Type II AGN in the sample
are discussed in section 4.4. The 215 galaxies that remain are
hereafter referred to as ultraviolet luminous galaxies (UVLGs). These
galaxies span the redshift range from $z=0.053$ to $z=0.3$.

A large number of galaxy parameters derived from the SDSS spectra are
available in the value-added catalogs produced by the SDSS
collaboration. These catalogs are available at the SDSS website at the
Max Planck Institute (http://www.mpa-garching.mpg.de/SDSS). From these
catalogs we use the emission line fluxes, widths, and derived
metallicities. For more information on the derivation of these
parameters, see \citet{k03a,k03b,k03c}, \citet{b04}, and
\citet{trem04}. These catalogs do not include metallicities for
galaxies with an AGN contribution since this can strongly affect the
line strengths, so only a subset of our sample have metallicity
determinations. In addition, some galaxies have poor line flux
measurements because the emission lines in the fiber aperture are weak
or non-existent, {\it e.g.,} in galaxies with no star formation in the
central region of the galaxy. Thus line flux measurements exist for
only a subset of our sample.

\subsection{Spectral Energy Distribution Modeling}	

To gain further information about the properties of the galaxies in
our sample, we compared the observed optical and UV properties of our
sample to a library of model spectral energy distributions (SEDs),
following \citet{salim05}. This was done by first constructing the
broadband optical and UV SEDs from the SDSS and \galex\
magnitudes. Each observed SED was then compared to an extensive
library of SEDs generated by the \citet{bc03} population synthesis
code. Each model galaxy is based on a star formation history composed
of an exponentially declining star formation rate (SFR) with
superimposed bursts of star formation, and includes the effects of
attenuation by dust (see Charlot \& Fall 2000). The library contains
$10^5$ models at each of five evenly spaced redshifts from $z=0.05$ to
$z=0.25$, and the grid of models was constructed to span the likely
range of star formation histories.

The goodness of fit for a given model to an observed SED is then
translated to a probability that the parameters for that model apply
to the galaxy. Thus the parameters of the best fitting model will have
the highest probability, and a probability distribution can be
constructed for the entire library at the appropriate redshift. From
this the median and $95$\% confidence limits on each parameter can be
determined. This was done for a list of parameters including stellar
mass and star formation rate over a range of timescales. For more
information on the SED fitting process, see \citet{salim05}. In this
paper the stellar masses and star formation rates were determined
through SED fitting.

\section{Properties of the GR1/DR3 Galaxy Sample}	

The \galex-SDSS matched catalog provides a valuable resource for
studying the UV-optical properties of star-forming galaxies in the
local universe. In a future paper we will report on the analysis of
the entire galaxy sample. Here we concentrate on the relationship
between UVLGs and the broader galaxy population.

The galaxy sample considered here should be nearly devoid of
unobscured (type I) AGN, so the dominant source of the UV light
detected by \galex\ is massive stars. The UV luminosity of a galaxy
therefore traces the total amount of star formation in that galaxy
over the past $10^8$ years \citep{cm05}.  We also have measurements of
the sizes of these galaxies. Most of these galaxies are only
marginally resolved in the \galex\ images, so we used the half-light
radii measured on the higher-resolution SDSS images. The SDSS $u$-band
was chosen because it is the closest in wavelength to the \galex\
bands and therefore the most likely to reflect the true spatial extent
of star-formation. In most cases we use the scale length from the
seeing-corrected exponential model fit calculated by the SDSS pipeline
as the half-light radius. However, for well-resolved galaxies we found
that the seeing-corrected radius derived from the exponential model
fits is systematically {\it larger} than the directly observed
half-light radius. We found that this occurs for galaxies with
half-light radii larger than about 2.2$^{\prime\prime}$. We thus use
the observed $u$-band half-light radius as $r_{50,u}$ for galaxies
larger than 2.2$^{\prime\prime}$, and the seeing-corrected scale
length as $r_{50,u}$ for galaxies smaller than
2.2$^{\prime\prime}$. We can then calculate the effective surface
brightness by dividing one-half of the luminosity by the area of the
galaxy enclosed by the half-light radius ($I_{1530}=L_{1530}/2\pi
r^2_{50,u}$).

\begin{figure}
\epsscale{1.0}
\plotone{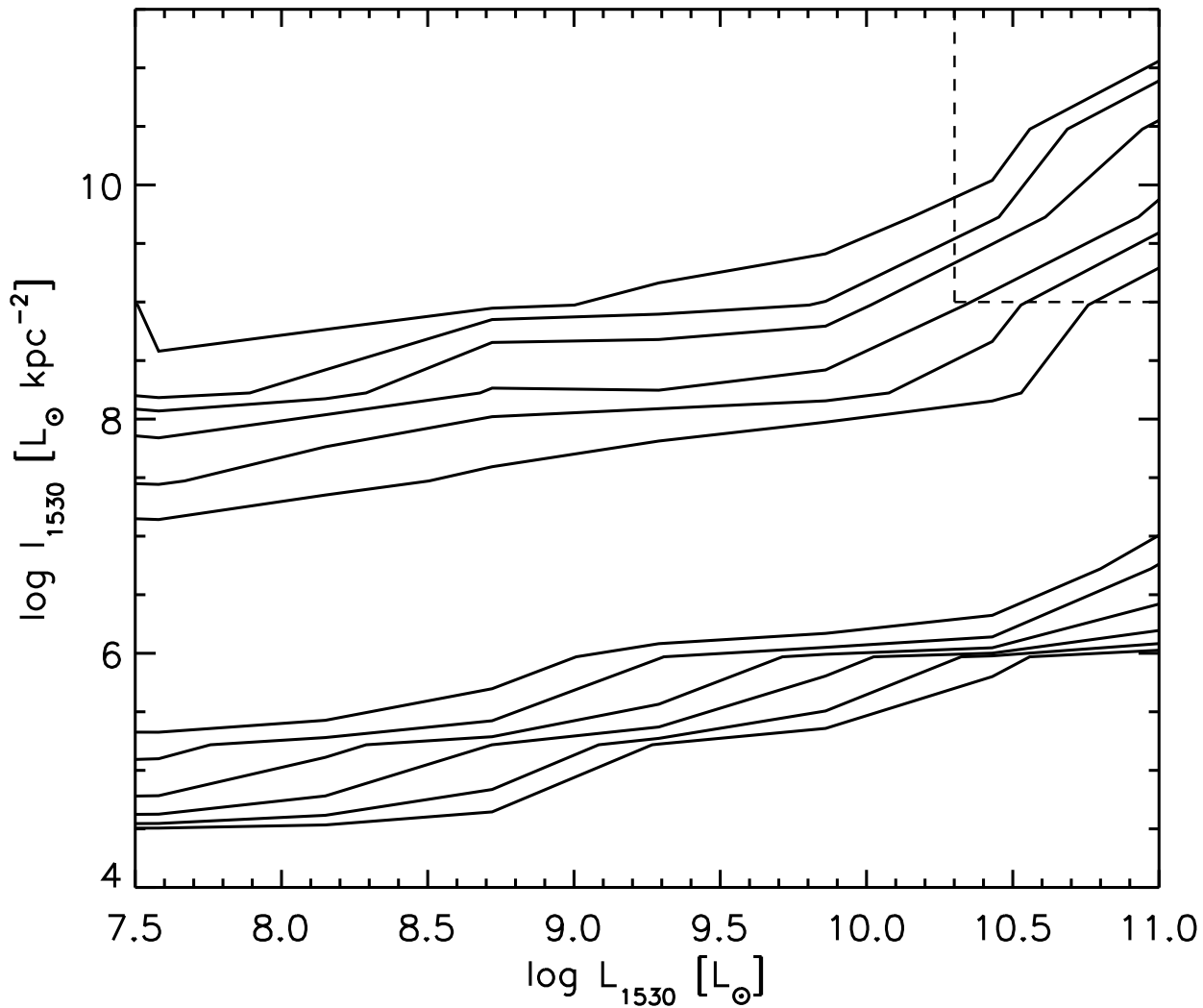}
\caption{
Normalized contour plot of the FUV surface brightness versus FUV
luminosity for 18463 galaxies in the GR1/DR3 sample with FUV
detections. The luminosity bins are normalized to have the same number
of galaxies in each bin. Each pair of contours represents a factor of
two increase in the enclosed fraction of galaxies in the luminosity bin that
have surface brightness in a given range, with the central pair of
contours enclosing 84\% of the galaxies in the luminosity bin, and the
outer pair of contours enclosing 99.5\% of the galaxies. FUV
luminosity ($L_{1530}$) is defined as $\lambda P_{\lambda}$ at
1530~\AA (observed wavelength). FUV surface brightness is defined as
$I_{1530}=L_{1530}/(2\pi r_{50,u}^2)$, where $r_{50,u}$ is the SDSS
$u$-band half-light radius (corrected for seeing).  The data points in
this figure have been corrected for Galactic foreground extinction,
but not for internal extinction. The dashed line shows the region
typically populated by LBGs.}
\label{sbfig}
\end{figure}

Figure~\ref{sbfig} shows a normalized contour plot of the FUV
surface brightness versus FUV luminosity for the 18463 galaxies in the
GR1/DR3 sample that were detected in the FUV images. The luminosity
bins were normalized to have the same number of galaxies in each bin,
thus clarifying the dependence of surface brightness on luminosity by
removing the effects of having a smaller number of galaxies at the low
and high luminosity ends of the distribution.

The plot shows a well-defined trend of slightly increasing surface
brightness with increasing luminosity over the entire luminosity
range. However, at the high luminosities corresponding to the UVLGs
there is an anomalous population of galaxies that defy the general
trend by having a much higher surface brightness than would be
expected given their luminosity. These galaxies have
$I_{1530}\ge10^{8}$~L$_{\odot}$~kpc$^{-2}$. Only among the UVLGs are
galaxies with the highest surface brightnesses
($I_{1530}\ge10^{9}$~L$_{\odot}$~kpc$^{-2}$) relatively common.

\begin{figure}
\epsscale{1.0}
\plotone{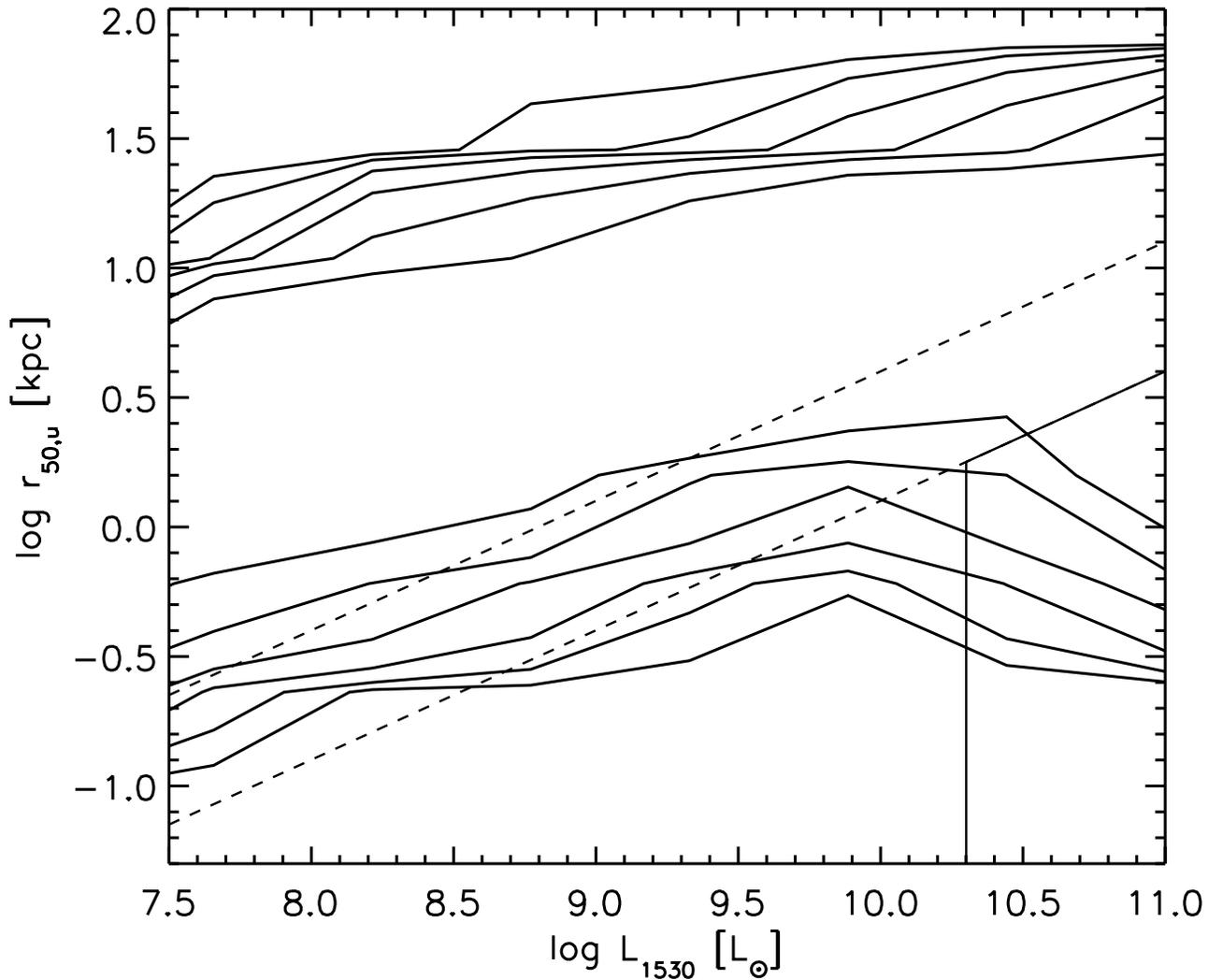}
\caption{Normalized contour plot of the $u$-band half-light radius versus 
FUV luminosity for 18463 galaxies in the GR1/DR3 sample that have
FUV detections. The luminosity bins are normalized to have the same
number of galaxies in each bin. Each pair of contours represents a factor of
two increase in the enclosed fraction of galaxies in the luminosity bin that
have surface brightness in a given range, with the central pair of
contours enclosing 84\% of the galaxies in the luminosity bin, and the
outer pair of contours enclosing 99.5\% of the galaxies. The SDSS $u$-band half-light
radius is derived from an exponential model fit and includes a seeing
correction. The dashed lines denote FUV surface brightness levels of
$I_{1530}=10^8$~L$_{\odot}$~kpc$^{-2}$ (upper line) and
$I_{1530}=10^9$~L$_{\odot}$~kpc$^{-2}$ (lower line). The solid line
shows the region typically populated by LBGs. The data points in this
figure have been corrected for Galactic foreground extinction, but not
for internal extinction.}
\label{sizefig}
\end{figure}

Figure~\ref{sizefig} shows the dependence of half-light radius on
luminosity. The surface brightness parameter shown in
Figure~\ref{sbfig} depends on the half-light radius, so
Figure~\ref{sizefig} is an alternative representation of
Figure~\ref{sbfig}. Over most of the range in luminosity the radius
increases with increasing luminosity. Above
$L_{1530}>10^{10}$~L$_{\odot}$ there is a group of galaxies that do
not obey this trend in the sense that they are too small for their
luminosity. The dashed lines in Figure~\ref{sizefig} are lines of
constant $I_{1530}$, and the galaxies responsible for this deviation
from the trend have $I_{1530}>10^9$~L$_{\odot}$~kpc$^{-2}$.

These two figures show that in general galaxies with higher UV
luminosity are larger and have somewhat higher surface brightness than
their less luminous counterparts. At high luminosities, however, some
galaxies behave differently. They have small radii, but this is more
than compensated by their increased surface brightness to put them
among the most UV luminous galaxies in the sample. This is a strong
indication that these UVLGs are distinct from the general galaxy
population. By contrast, UVLGs with lower surface brightness do not
distinguish themselves from the full sample of galaxies except by
their luminosity. They appear to be the largest and therefore most
luminous normal galaxies.

\begin{figure}
\epsscale{1.0}
\plotone{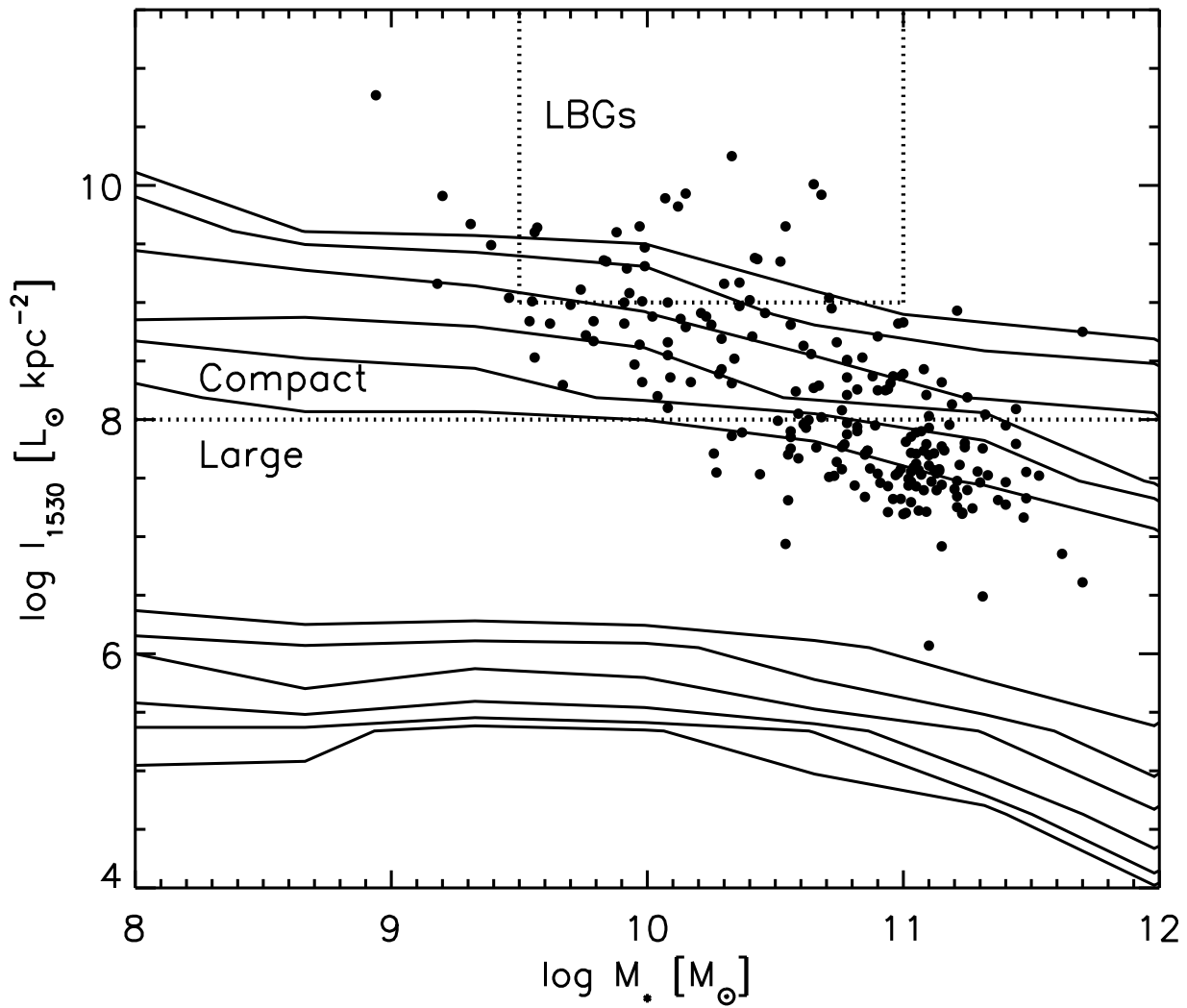}
\caption{
Normalized contour plot of the FUV surface brightness versus stellar
mass for the 18463 galaxies in the GR1/DR3 sample that have FUV
detections. The stellar mass bins are normalized to have the same
number of galaxies in each bin. Each pair of contours represents a
factor of two increase in the enclosed fraction of galaxies in the
mass bin that have surface brightness in a given range, with the
central pair of contours enclosing 84\% of the galaxies in the mass
bin, and the outer pair of contours enclosing 99.5\% of the
galaxies. The individual galaxies in the UVLG sample are shown as
points in the plot. The boundaries of mass and surface brightness for
typical LBGs are also shown in the figure, as is the boundary between
large and compact UVLGs.}
\label{massfig}
\end{figure}

Figure~\ref{massfig} shows how the UV surface brightness changes with
stellar mass, as determined from the SED model fitting
\citep{salim05}. In general the stellar mass determined in this manner
agrees to within a factor of 2 with the dynamical mass calculated as
$M_{dyn} = 3.4\sigma^2_{gas}r_{50,u}/G$, where $\sigma_{gas}$ is the
standard deviation of the gas velocity measured from the emission
lines. The coefficient 3.4 was taken from \cite{erb06a} and represents
a realistic estimate of the mass distribution for LBGs. The UV surface
brightness is relatively constant over a wide range of stellar masses,
and then slowly falls above a mass of $10^{10.5}$~M$_{\odot}$. This
implies that the more massive galaxies have correspondingly larger
sizes over which the young stellar population is distributed. The drop in
surface brightness above $10^{10.5}$~M$_{\odot}$ may be related to
the relatively abrupt transition in the galaxy population at this
mass-scale between young disk-dominated galaxies and old
bulge-dominated ones (e.g. Kauffmann et al. 2003b).

The points shown in Figure~\ref{massfig} are the locations of
individual UVLGs. Unlike the galaxy population as-a-whole, the UVLGs
show a clear inverse correlation between surface brightness and
mass. We have already pointed out that this fact indicates that the
more massive UVLGs owe their large luminosities to their large
mass. The less massive UVLGs have high UV surface brightnesses
indicative of intense star-formation. Figure~\ref{massfig} shows how
the UVLGs relate to the rest of the sample in terms of mass (but keep
in mind that the contours are normalized by the number of galaxies in
each mass bin). The UVLGs with $I_{1530}<10^8$~L$_{\odot}$~kpc$^{-2}$
are among the most massive star-forming galaxies in the GR1/DR3
sample, with log $M_*\ge10.5$~M$_{\odot}$. While these are the lower
surface brightness component of the UVLG sample, they are still
somewhat offset toward higher surface brightness than the full sample
(they are not low surface brightness galaxies). In the main, their
properties appear to be similar to those of large, disk galaxies (they
are the extrema of the population).

The UVLGs with $I_{1530}>10^8$~L$_{\odot}$~kpc$^{-2}$ are generally
lower mass systems (log $M_*\le10.5$~M$_{\odot}$). They clearly stand
out from the full sample by having much higher surface brightness than
would be expected for normal galaxies of similar mass. This is even
more obvious for galaxies with $I_{1530}>10^9$~L$_{\odot}$~kpc$^{-2}$,
which would qualify as LBGs based on their FUV surface brightness.

Based on the analysis above, the UVLG population can be thought of as
two very different types of galaxies. The high-surface brightness
systems (``compact UVLGs'') have high star formation rates per unit
area and would be called starburst galaxies, while the low surface
brightness systems (``large UVLGs'') are large spirals, with high
rates of total star formation but low rates of star formation per unit
area. There is no clear transition from one population to the other,
but a surface brightness value of
$I_{1530}=10^8$~L$_{\odot}$~kpc$^{-2}$ serves as a useful boundary.
There are intermediate cases that do not fit cleanly into either
category. Figure~\ref{massfig} shows that this surface brightness
boundary corresponds to a stellar mass of roughly
$M_*=10^{10.5}$~M$_{\odot}$ (similar to the mass scale that divides
the bimodal galaxy population as-a-whole -- e.g. Kauffmann et
al. 2003b).  Using this criterion there are 110 large UVLGs and 105
compact UVLGs in the GR1/DR3 sample. These two diverse populations
were recognized in Paper I, but we can now place them firmly in the
context of the overall galaxy population.

Throughout the rest of this paper we will distinguish between large
and compact UVLGs. Note, however, that while the compact UVLGs have
the properties of intense starbursts, not all of them have FUV surface
brightnesses high enough to be considered typical LBGs, which
generally have $I_{1530}>10^9$~L$_{\odot}$~kpc$^{-2}$. We will
consider the compact UVLGs that meet this more stringent surface
brightness criterion as possible LBG analogs, and will refer to them
as ``supercompact UVLGs''. The GR1/DR3 sample contains 35 supercompact
UVLGs.

\section{Properties of Ultraviolet Luminous Galaxies}	

Since the UVLG sample was chosen based on an ultraviolet luminosity
criterion, they are all expected to have high star formation rates. As
in Paper I, the majority of UVLGs (83\%) have concentration parameters
$C<2.6$, where $C$ is defined as $R_{90}/R_{50}$, the ratio of the
radius containing 90\% of the Petrosian $r$-band luminosity to that
containing 50\%. These low concentration parameters are indicative of
disk systems, as expected for a sample of star-forming galaxies. Yet
as was made clear in the previous section, UVLGs span a wide range of
properties. In this section we explore the properties of UVLGs. The
properties of the 215 UVLGs are listed in Table~\ref{samptab}.

\subsection{Ultraviolet Surface Brightness}	

\begin{figure}
\epsscale{0.5}
\plotone{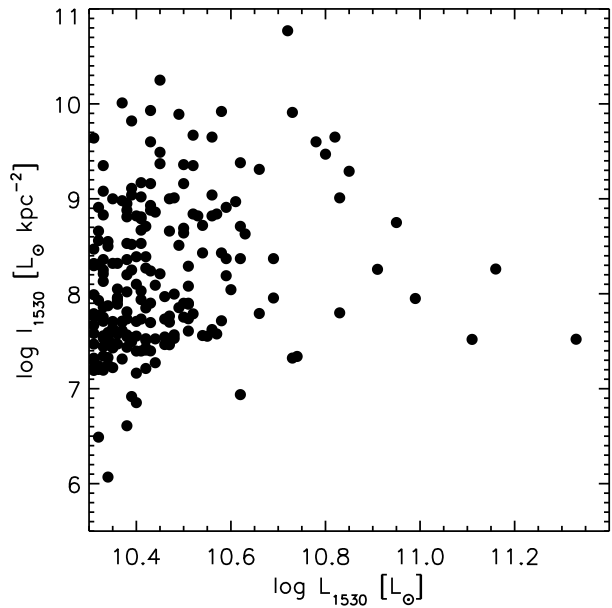}
\caption{
FUV surface brightness versus FUV luminosity for 215 UVLGs in the
GR1/DR3 sample. FUV luminosity ($L_{1530}$) is defined as $\lambda
P_{\lambda}$ at 1530~\AA. FUV surface brightness is defined as
$I_{1530}=L_{1530}/(2\pi r_{50,u}^2)$, where $r_{50,u}$ is the SDSS
$u$-band half-light radius.}
\label{lvifig}
\end{figure}

\begin{figure}
\epsscale{0.5}
\plotone{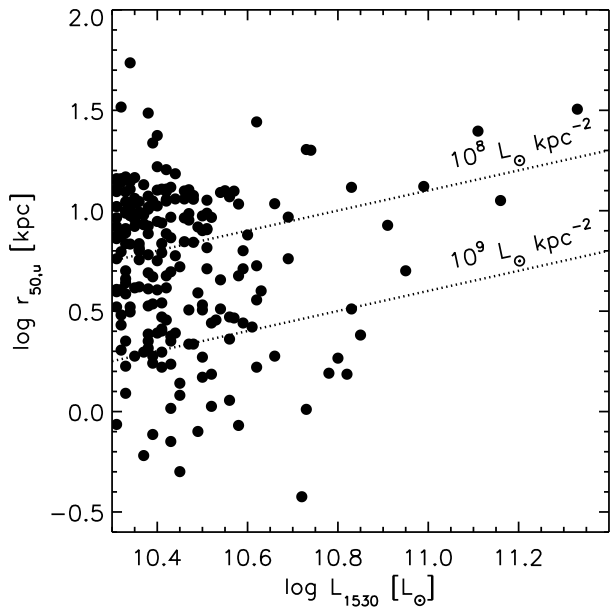}
\caption{
Half-light radius in the SDSS $u$-band versus FUV luminosity for 215
UVLGs in the GR1/DR3 sample. FUV luminosity ($L_{1530}$) is defined as
$\lambda P_{\lambda}$ at 1530~\AA. The upper dotted line shows a
constant surface brightness of $10^8$~L$_{\odot}$~kpc$^{-2}$, which is
our chosen boundary between large and compact UVLGs, and the lower
dotted line shows $I_{1530}=10^9$~L$_{\odot}$~kpc$^{-2}$, which is the
lower boundary of values seen in typical LBGs at $z=3$ \citep{g02}.}
\label{lvrfig}
\end{figure}

Figure~\ref{lvifig} plots the FUV surface brightness of the 215 UVLGs
against the FUV luminosity. The galaxies were chosen to be luminous,
but they span a wide range in surface brightness, and there is no
correlation between luminosity and surface brightness. This implies
that the UVLGs span a similarly large range of size. This is confirmed
in Figure~\ref{lvrfig}, which plots the luminosity against the
half-light radius. UVLGs range in half-light radius from less than a
kpc to $>20$~kpc. The dotted lines in Figure~\ref{lvrfig} show a
constant surface brightness of $10^8$~L$_{\odot}$~kpc$^{-2}$ and
$10^9$~L$_{\odot}$~kpc$^{-2}$, the latter being the lower limit seen
in LBGs at $z=3$ \citep{g02}. Only a fraction of the UVLGs have
surface brightnesses that rival those of LBGs, even though they all
have LBG-like luminosities.

The FUV surface brightness is related to the star formation intensity,
${\it i.e.,}$ the star formation rate per unit
area. Figures~\ref{lvifig} and \ref{lvrfig} show then that only a
subset of the UVLGs are luminous because they have high star formation
intensities. The rest owe their high luminosities to their large size,
{\it i.e.,} they have modest levels of star formation intensity spread
over a large area.

This is also apparent in Figure~\ref{massfig}, which shows more clearly
the correlation between surface brightness and stellar mass noted
above.  The UVLGs with low surface brightness are the most massive,
while the high surface brightness UVLGs are low-mass systems. The
typical mass and surface brightness range of LBGs is shown in the
figure \citep{shap01,pap01,g02}. Figures $3-5$ illustrate the fact
that UVLGs span a continuous range of properties, {\it i.e.,} there is
no clear demarcation between the large and compact samples. The
division of the samples at a surface brightness of
$I_{1530}>10^8$~L$_{\odot}$~kpc$^{-2}$ is an arbitrary boundary.

\subsection{Star Formation and Attenuation by Dust}	

\begin{figure}
\epsscale{0.5}
\plotone{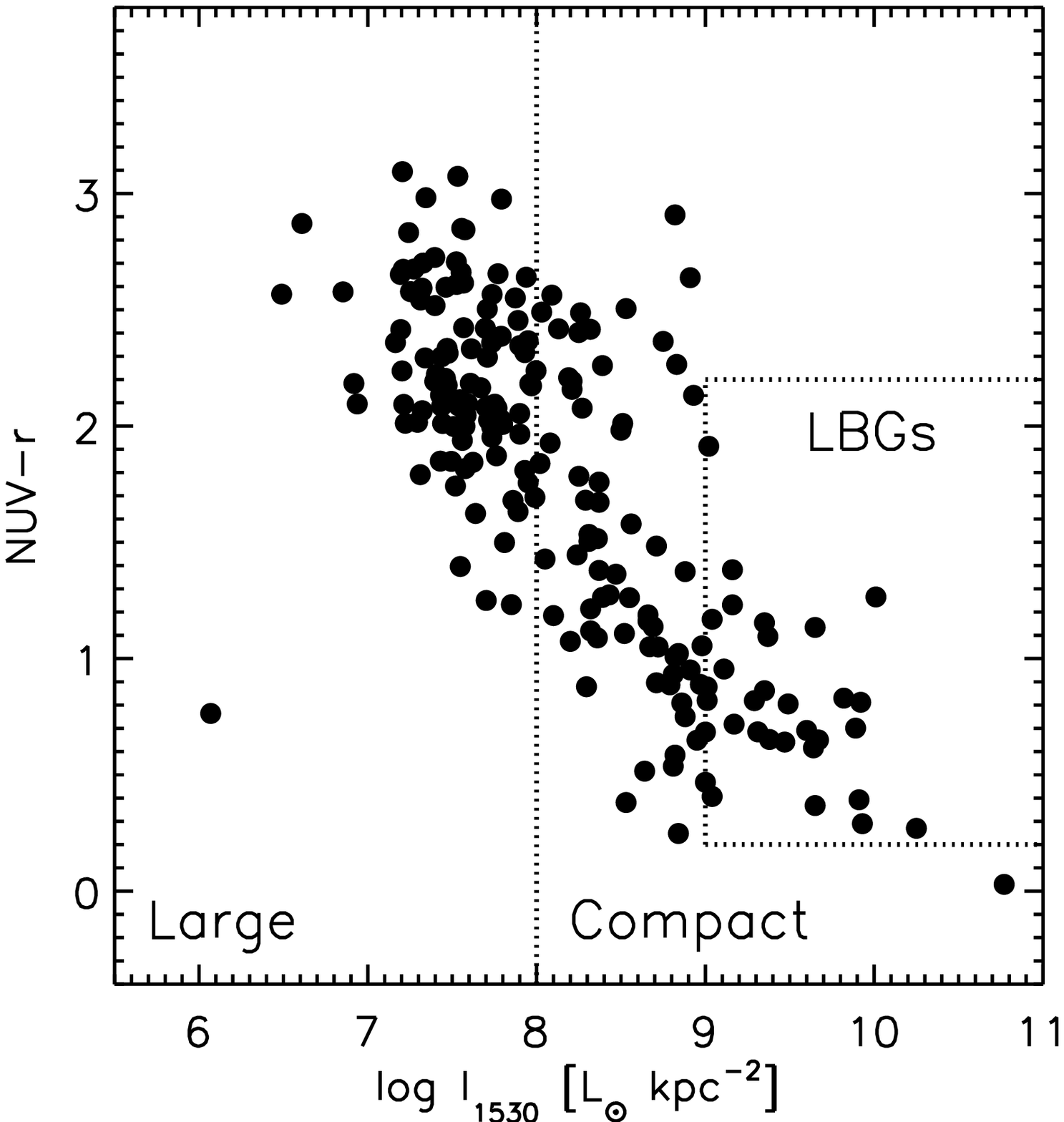}
\plotone{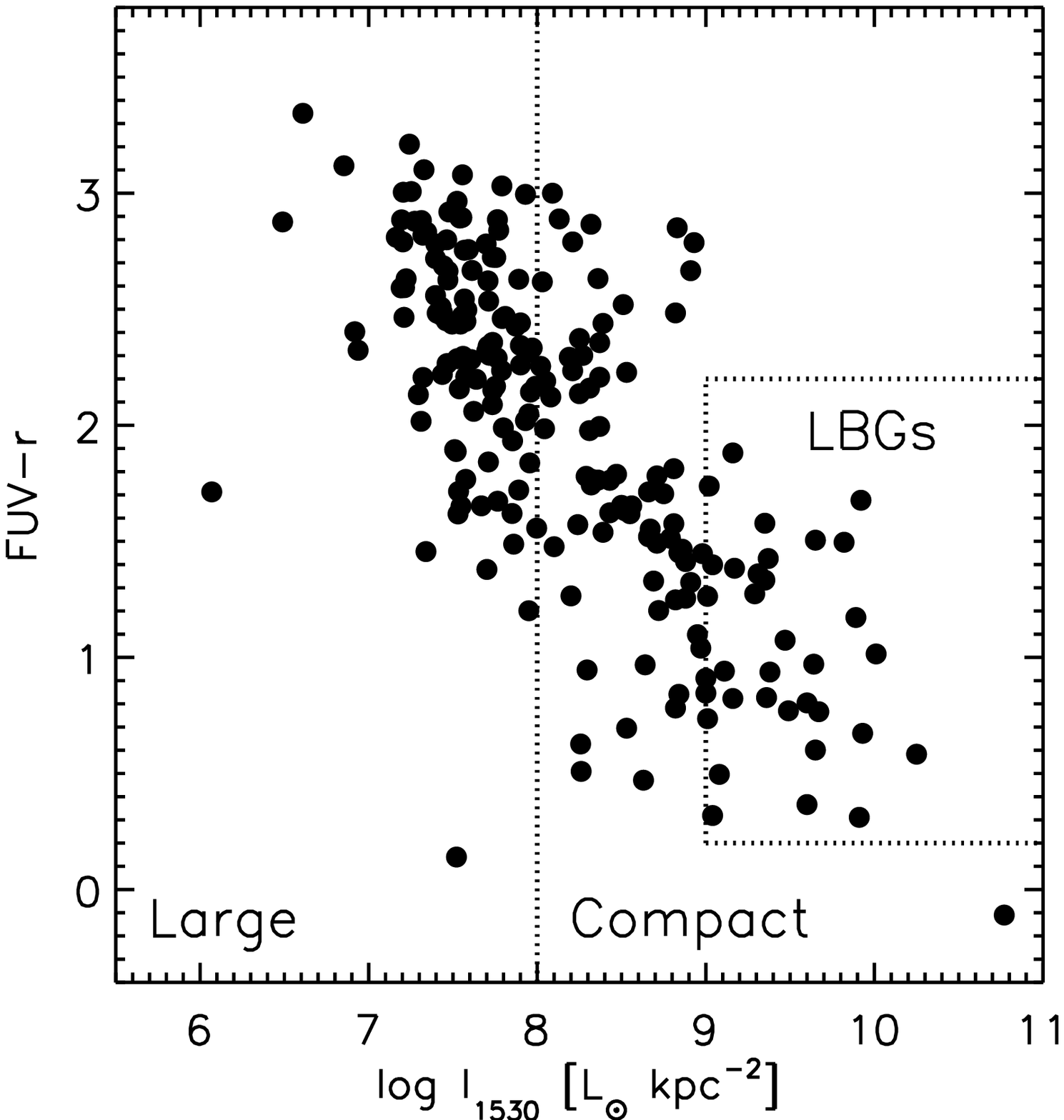}
\caption{{\it top:} 
FUV-$r$ color versus FUV surface brightness for 215 UVLGs in the
GR1/DR3 sample. FUV surface brightness is defined as
$I_{1530}=L_{1530}/(2\pi r_{50,u}^2)$, where $r_{50,u}$ is the SDSS
$u$-band half-light radius. The dotted line shows the boundaries of
FUV-$r$ color and surface brightness for typical LBGs. {\it bottom:}
NUV-$r$ color versus FUV surface brightness for 215 UVLGs in the
GR1/DR3 sample.}
\label{frvifig}
\end{figure}

Figure~\ref{frvifig} shows the FUV-$r$ color and NUV-$r$ color for the
215 UVLGs as a function of surface brightness. Both colors are
well-correlated with surface brightness, with the brightest galaxies
having the bluest color. This agrees with the idea that the UV-optical
colors are sensitive to the ratio of current to past star
formation. \citet{salim05} showed the NUV-$r$ in particular is a good
tracer of the star formation rate parameter $b$, which is the current
SFR divided by the past-average SFR.  The blue color of the high
surface brightness UVLGs can be understood if they are undergoing
intense starbursts which are much more significant than the past
average rate of star formation. The FUV surface brightness appears to
be a good indicator of star formation intensity for UV-selected
galaxies. The typical colors of LBGs are also indicated in the plot
\citep{shap01,pap01,g02}.

\begin{figure}
\epsscale{0.43}
\plotone{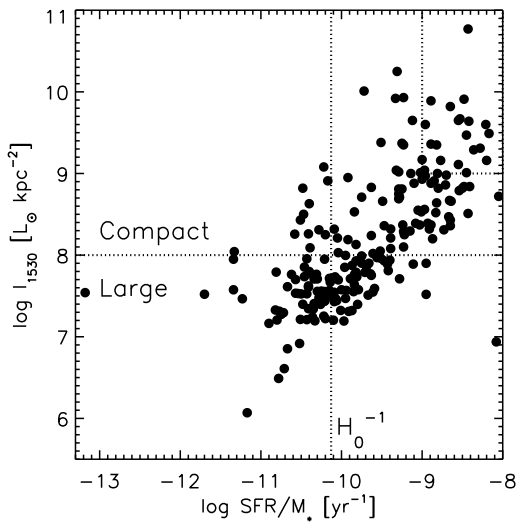}
\caption{
FUV surface brightness versus specific star formation rate
(extinction-corrected star formation rate divided by stellar mass) for
215 UVLGs in the GR1/DR3 sample. The stellar mass and star formation
rate were determined via SED fitting. FUV surface brightness is
defined as $I_{1530}=L_{1530}/(2\pi r_{50,u}^2)$, where $r_{50,u}$ is
the SDSS $u$-band half-light radius. The boundaries of specific SFR
and surface brightness for typical LBGs are marked (the small region
in the upper right outlined by a dotted line), as is the value of the
Hubble time for our chosen cosmology.}
\label{sfrvifig}
\end{figure}

Figure~\ref{sfrvifig} shows the FUV surface brightness versus the
specific star formation rate (star formation rate normalized by
stellar mass). The extinction-corrected star formation rates were
determined by SED model fitting. We have also calculated star
formation rates using the \ha\ luminosity in the SDSS spectra using
the recipe given in \cite{k98}, these values generally agree within a
factor of 2, which is quite good considering that the H-alpha
measurements were taken through 3$^{\prime\prime}$ fibers. The
specific star formation rate relates the current to past star
formation, and the inverse of this quantity is the ``galaxy building
time,'' the time it would take to build up the current stellar mass at
the current SFR. The specific star formation rate is clearly
correlated with the FUV surface brightness, with the high surface
brightness systems generally have the highest specific SFRs and short
building times, indicating that these are starburst systems. The large
UVLGs have galaxy building times of roughly a Hubble time, as expected
for a galaxy that has been built up over the age of the universe at a
constant or slowly varying rate of star formation. Galaxy building
times of less than 1 Gyr are typical for LBGs
\citep{shap01,pap01,g02}, and the high surface brightness UVLGs
overlap this range. However some of the high surface brightness
systems have lower specific SFRs and longer building times than
typical LBGs, which suggests that they have had significant star
formation prior to the current burst. $Spitzer$ imaging of LBGs
indicates that they do not have significant populations of older stars
\citep{ba04}, suggesting that LBGs are undergoing their first major burst
of star formation. If this is the case, then these systems with longer
building times may not be true analogs for LBGs. However, there is a
significant fraction of the high-surface brightness systems that fall
in the boundaries set by the LBGs, and these may be excellent analogs.

\begin{figure}
\epsscale{0.43}
\plotone{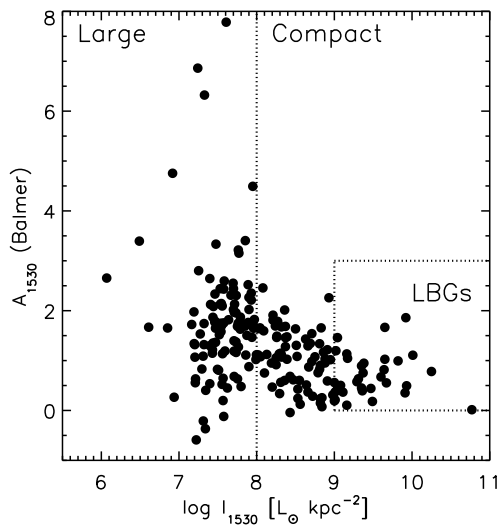}
\caption{
Attenuation of the FUV continuum versus FUV surface brightness for the
210 UVLGs in the GR1/DR3 sample with good line flux measurements. FUV
surface brightness is defined as $I_{1530}=L_{1530}/(2\pi
r_{50,u}^2)$, where $r_{50,u}$ is the SDSS $u$-band half-light
radius. The FUV attenuation was determined using the Balmer decrement
in the SDSS spectra and the \cite{calzetti01} starburst attenuation
law, in which $E(B-V)_{continuum}=0.44E(B-V)_{lines}$. Four large
UVLGs are not shown because have no emission lines in the SDSS fiber
spectra, and one supercompact UVLG has an apparent anomaly in the
spectrum near H$\alpha$. The dotted line shows the boundaries of FUV
attenuation and surface brightness for typical LBGs.}
\label{avifig}
\end{figure}

Figure~\ref{avifig} shows the FUV attenuation as a function of surface
brightness. The FUV attenuation was determined using the Balmer
decrement and the \cite{calzetti01} starburst attenuation law. The
compact UVLGs are in the range $A_{1530}\le2$, indicating that a
relatively large fraction ($>10\%$) of the UV light escapes. The large
UVLGs are mostly in the range from 0 to $>4$ magnitudes of
attenuation. Figure~\ref{avifig} shows that compact UVLGs have a lower
amount of attenuation on average than do large UVLGs, and the higher
surface brightness compact UVLGs have still lower average extinction
values.  We note that this method of determining attenuation values
uses the SDSS fiber spectra and may be sensitive to aperture
effects. The fibers capture most of the light from the compact UVLGs,
so this should only be an issue for the large UVLGs, where we are
measuring the attenuation in the central core of the
galaxy. Figure~\ref{avifig} also shows the typical range of FUV
attenuation seen in LBGs \citep{shap01,pap01}.

\subsection{Metallicity}	

\begin{figure}
\epsscale{1.0}
\plottwo{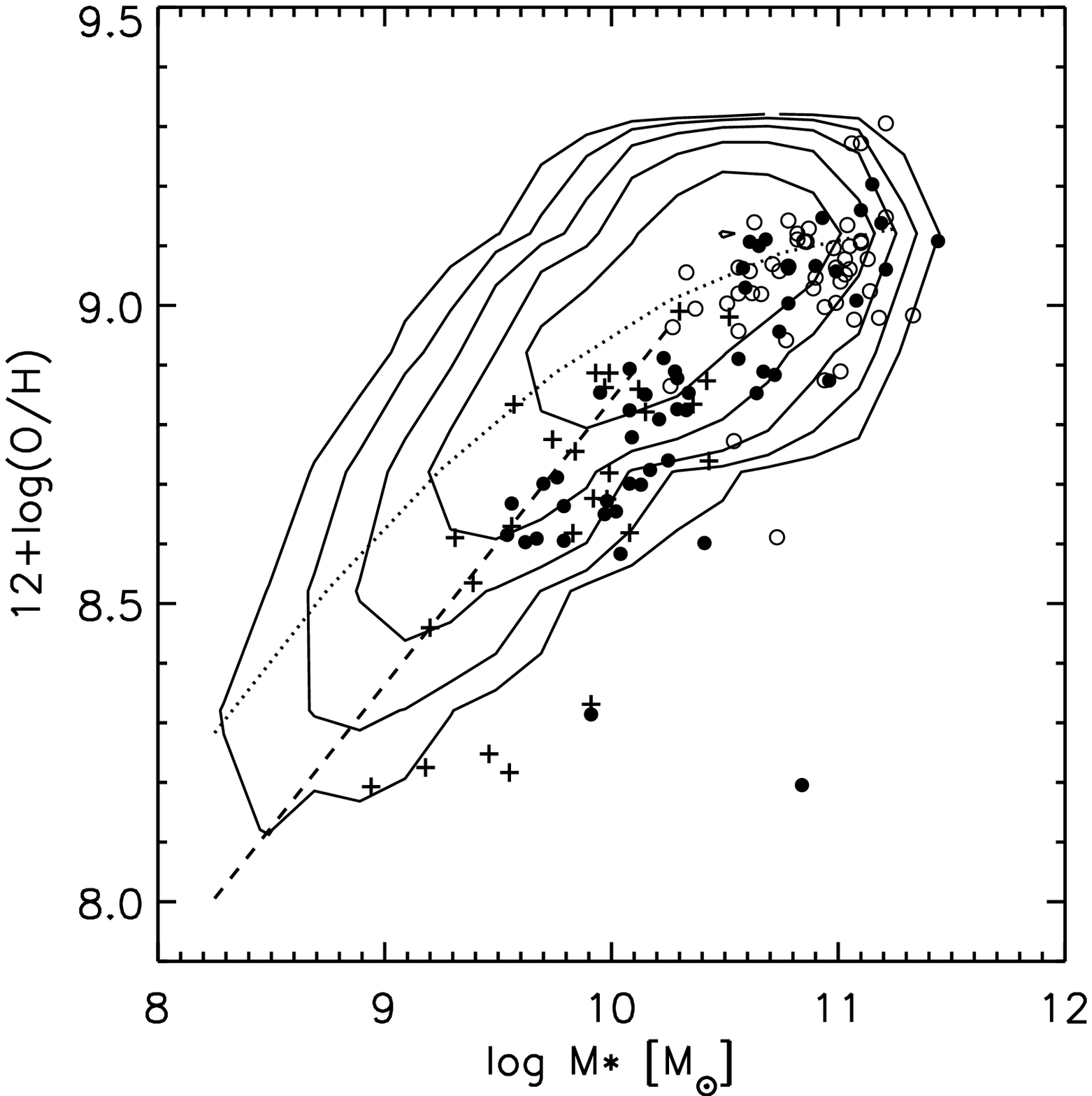}{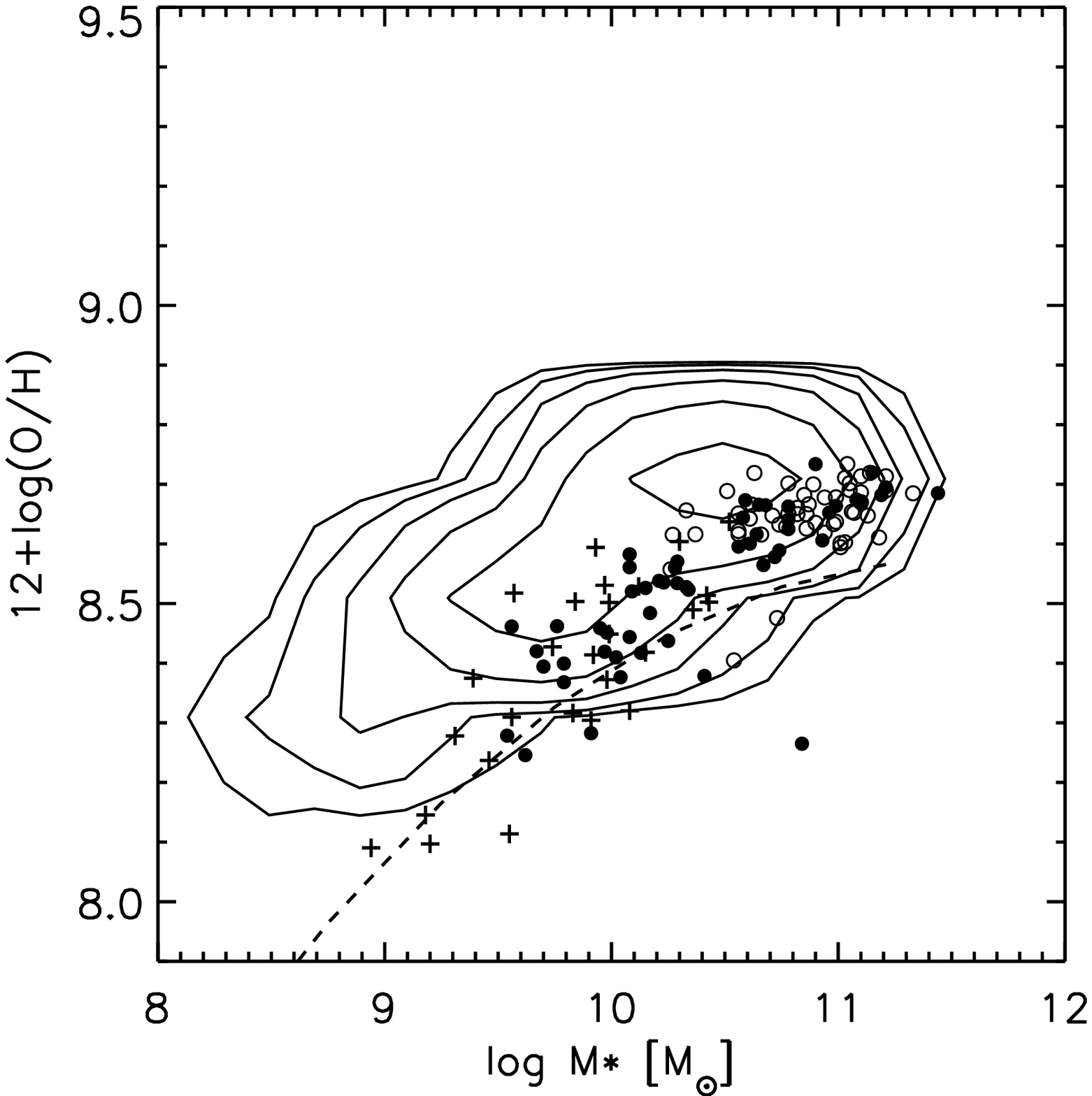}
\caption{
{\it Left:} The dependence of metallicity upon stellar mass. The O/H
estimates were derived from the SDSS spectra using the
technique described in \cite{trem04}, and the stellar mass was
determined via SED fitting \citep{salim05}. The contours represent
10002 galaxies in the GR1/DR3 sample which have good metallicity
measurements. The individual galaxies in the UVLG sample are shown as
points in the plot. Open circles are UVLGs with
$I_{1530}<10^8$~$L_{\odot}$~kpc$^{-2}$ (large UVLGs), and filled
circles are UVLGs with
$10^8$~$L_{\odot}$~kpc$^{-2}<I_{1530}<10^8$~$L_{\odot}$~kpc$^{-2}$
(compact UVLGs). UVLGs with $I_{1530}>10^9$~$L_{\odot}$~kpc$^{-2}$
(supercompact UVLGs) are denoted as crosses. The dotted line is the
best fit to the \cite{trem04} sample, and the dashed line is the fit
from \cite{ss05}. {\it Right:} Same as the left panel, but using the
$N2$ method as calibrated by
\cite{pp04}. The dashed line is from
\cite{erb06a}, which is the \cite{trem04} fit displaced downward by
0.56 dex. }
\label{allmetfig}
\end{figure}

Figure~\ref{allmetfig} shows two determinations of the mass {\it vs.} 
metallicity relation for the entire GR1/DR3 sample. The left panel
shows the relation determined via the method of \cite{trem04}. This
method makes use of a grid of models, which combine the \cite{bc03}
population synthesis code with photo-ionization models of \ion{H}{2}
regions. The methodology is described in detail in
\cite{cl01}. Metallicities are constrained by fitting all the strong
emission lines in the SDSS spectra. Only 129 of the UVLGs have
metallicity determinations from the SDSS spectra, because galaxies
with an AGN contribution to their spectra were excluded. The
well-known mass-metallicity relation \citep{trem04} is apparent in
Figure~\ref{allmetfig}, and the best fit to the Tremonti et al. sample
is shown as a dotted line.  The UVLGs also show a relation between
mass and metallicity, but it is offset from the general sample. At
high masses ($>10^{10.5}$ M$_{\odot}$) most of the UVLGs have
metallicities similar to those of normal galaxies. Most of these
objects are the large UVLGs, which we have argued are just the
UV-bright tail of the population of normal high-mass star forming
galaxies. Their relatively normal metallicities support this idea.  At
lower masses (where the sample is primarily the compact UVLGs), the
slope of the mass-metallicity relation for the UVLGs is significantly
steeper than that of the overall galaxy population. In the mass range
$\sim10^9$ to $10^{10}$~M$_{\odot}$, the compact UVLGs have
metallicities a factor of two to three lower than normal galaxies of
the same mass.

Selection on the basis of high UV luminosity could clearly bias the
sample against dusty objects, and the dust/gas ratio will be larger
for higher metallicity. Thus, selection effects could make a UV-bright
sample have systematically lower metallicity. While such an effect may
be present, it does not explain the mass-dependence of the offset
between the mass-metallicity relation for the sample as-a-whole, and
that of the UVLGs.

It is very interesting to compare the mass-metallicity relation for
the UVLGs to what is found for UV-bright galaxies at higher redshifts.
The form of the mass-metallicity relation for the UVLGs is similar to
that found by \citet{ss05}, who investigated a sample of star-forming
galaxies at $z=0.7$. Their relation is shown in the left panel of
Figure~\ref{allmetfig}.  They interpreted the change in the form of
the mass-metallicity relation from $z\sim0.1$ to $\sim0.7$ in terms
of the ``down-sizing'' of the galaxy population. Massive galaxies have
essentially the same metallicity at $z \sim$ 0.1 and $\sim$0.7 because
this population has already come near the end-point of its evolution
by $z\sim0.7$. The strong chemical evolution at late times seen in
the low mass galaxies is because they are still converting significant
mass from gas into stars at the present time. Applied to the compact
UVLGs, this would suggest that they are relatively unevolved compared
to typical galaxies of the same mass. Alternatively, perhaps the
compact UVLGs are UV-bright because they are experiencing a burst of
star formation triggered by the infall of metal-poor gas.

\cite{erb06a} have measured the mass-metallicity relation for LBGs
at $z\sim2$. They find that this relation has evolved considerably
compared to the local universe, with LBGs having systematically lower
metallicities than present-day galaxies by a factor of about two {\it
for all masses}. Erb et al. estimate metallicity using the $N2$
method, which uses the [\ion{N}{2}]/H$\alpha$ ratio, as calibrated by
\cite{pp04}. The resulting metallicities are known to be
systematically lower than those obtained using the \cite{trem04}
method. To fairly compare the UVLGs and LBGs, we also plot our
mass-metallicity relation based on the \cite{pp04} method (right panel
of Figure~\ref{allmetfig}). While the shape of the relation changes,
the offset of the compact UVLGs from the rest of the galaxy population
at low mass remains. The dashed line shows the relation for LBGs from
\cite{erb06a}. At low masses ($<10^{10.5}$ M$_{\odot}$) the compact
UVLGs and LBGs have similar metallicity. At higher masses the UVLGs
are more metal-rich (including the compact UVLGs). Erb et
al. interpret the relation for the LBGs in terms of the loss of metals
by supernova-driven winds that is occurring at all masses. Results by
Shapley et al. (2005) suggest this may also be true at redshifts as
low as ~1 to 1.5 \citep{scmb05}. This contrasts with the
mass-dependent loss of metals inferred for the local galaxy population
\citep{trem04}. The form of the mass-metallicity relation for the
UVLGs suggests mass-dependent metal loss. We will consider this idea
in detail in a future paper.

\begin{figure}
\epsscale{0.45}
\plotone{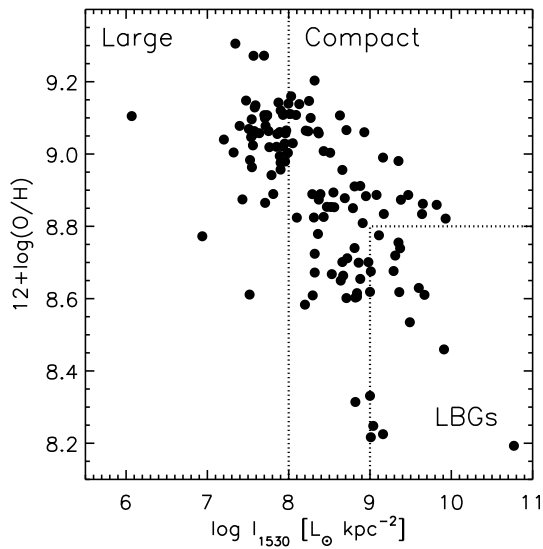}
\caption{
FUV surface brightness versus metallicity for 129 UVLGs in the GR1/DR3
sample which have good metallicity measurements (galaxies with an AGN
contribution to their spectra were removed). FUV surface brightness is
defined as $I_{1530}=L_{1530}/(2\pi r_{50,u}^2)$, where $r_{50,u}$ is
the SDSS $u$-band half-light radius. The O/H estimate is derived from
the SDSS spectra (Tremonti et al. 2004). The dotted line shows the
boundaries of 12+log(O/H) and surface brightness for typical LBGs.}
\label{metfig}
\end{figure}

Figure~\ref{metfig} shows the metallicity of the ionized gas,
determined from the SDSS spectra \citep{trem04}, in UVLGs as a
function of FUV surface brightness. There is a
clear trend of increasing metallicity with decreasing surface
brightness. Given the connection between surface brightness and mass
previously discussed (see Figure~\ref{massfig}), this is likely a
reflection of the mass-metallicity correlation. The high surface
brightness galaxies have the lowest metallicity, and this metallicity
is typical of that found in LBGs \citep{shap04,erb06a}.

\subsection{Active Galactic Nuclei}	

\begin{figure}
\epsscale{0.48}
\plotone{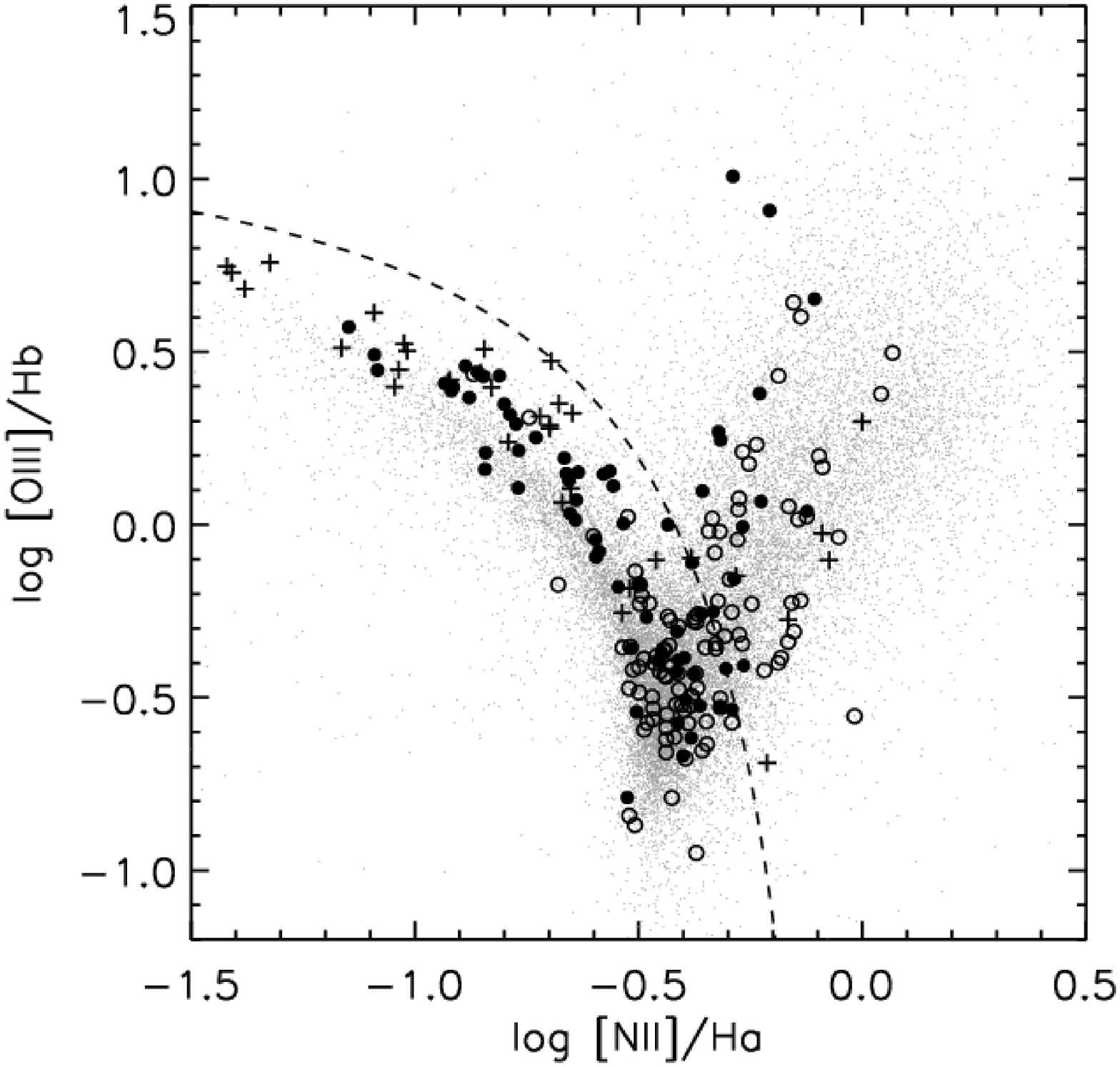}
\caption{
A BPT diagram \citep{bpt81} for the 23400 galaxies with good line flux
measurements in the DR3/GR1 sample. The dashed curve
shows the demarcation between starburst and AGN defined by
\citet{k03c}. The large circles and crosses are UVLGs. There are 106
large UVLGs (open circles), 70 compact UVLGs (filled circles), and 34
supercompact UVLGs (crosses) shown in this plot. Four large UVLGs are
not shown because have no emission lines in the SDSS fiber spectra,
and one supercompact UVLG has an apparent anomaly in the spectrum near
H$\alpha$. }
\label{bpt}
\end{figure}

Figure~\ref{bpt} is a diagnostic diagram that uses line ratios to
differentiate between active galactic nuclei (AGN) and
star-formation-dominated systems \citep{bpt81}. The dashed line in the
figure shows the line of demarcation determined by \citet{k03a} for
SDSS galaxies, where galaxies below the line are dominated by star
formation. The UVLGs are shown in the figure as large circles, with
filled circles denoting UVLGs with
$I_{1530}>10^8$~L$_{\odot}$~kpc$^{-2}$ and crosses denoting UVLGs with
$I_{1530}>10^9$~L$_{\odot}$~kpc$^{-2}$. Of the 104 compact (and
supercompact) UVLGs in this plot, 22 are classified as AGN or
transition objects, roughly 21\%, compared to 33\% of the entire
sample in the compact UVLG mass range ($9.0$~M$_{\odot} \le $~log $M_*
\le 11.0$~M$_{\odot}$). About 34\% of
the large UVLGs in this diagram
($I_{1530}<10^8$~L$_{\odot}$~kpc$^{-2}$) are classified as AGN (36 out
of 106), while for the entire sample in this mass range
($10.3$~M$_{\odot} \le $~log $M_* \le 11.7$~M$_{\odot}$) the fraction is
54\%.  This difference may reflect the fact that it
can be more difficult to recognize a type II AGN in a starburst due to
the strong emission lines produced by star formation. The majority of UVLGs have the line ratios of normal
star-forming galaxies. Note that type I AGN have been removed from the
GR1/DR3 sample because the AGN would contribute significantly to the
UV luminosity.

\section{Discussion}	

\subsection{Large and Compact UVLGs}	

The most luminous galaxies in the ultraviolet
($L_{1530}>2\times10^{10}$~L$_{\odot}$) are a diverse population
spanning a wide range of properties. Most of these properties are
correlated at some level with the FUV surface brightness, which
reflects the star formation rate per unit area. It is informative to
use the surface brightness distribution to divide the UVLGs into two
groups, ``large'' and ``compact''.

The large UVLGs at the low surface brightness end are very massive
spirals. They have stellar masses of $M_*>10^{10.5}$~M$_{\odot}$,
comparable to massive spirals in the local universe. They have
half-light radii of 5 to 30~kpc. They are at the extreme end of the UV
luminosity distribution, an indication that they have high global
star formation rates. However, this high luminosity is a result of
relatively modest star formation intensity spread out over a large
area. They seem to share many of the characteristics of normal
spirals but extend this population to high UV luminosity. By
selecting the most luminous galaxies we have found the tail of the
distribution of massive spirals.

The compact UVLGs at the high-surface brightness end are systems with
intense star formation in a relatively compact region. They are lower
mass systems ($M_*<10^{10.5}$~M$_{\odot}$) with half-light radii of 1
to 5~kpc. They have high specific star formation rates, suggesting
they may be experiencing their first major burst of star
formation. They are generally metal-poor compared to the large UVLGs.

Figures~\ref{sbfig} and \ref{sizefig} illustrate the basic differences
between large and compact UVLGs. These figures show that the large
UVLGs can be understood as the high-luminosity end of the distribution
of normal galaxies. They are extremely luminous, but this is mostly a
reflection of their large size. On the other hand, the compact UVLGs
deviate from the trends established by the full sample, in that they
are very luminous but not very large. Their high luminosity reflects
an extremely high surface brightness in a relatively small
region. This behavior sets them apart from the rest of the galaxies,
and suggests that compact UVLGs are a distinct population of galaxies.

\subsection{Local Analogs of Lyman Break Galaxies}	

One of the initial goals of this investigation was to find nearby
galaxies with properties most similar to LBGs. To do this, we have
defined a surface brightness cutoff at
$I_{1530}>10^9$~L$_{\odot}$~kpc$^{-2}$, which is the lower limit of UV
surface brightness seen in typical LBGs
\citep{g02}. Table~\ref{proptab} shows that the galaxies in this
sample (``supercompact UVLGs'') share all of the characteristics of
LBGs that are considered in this paper, including UV luminosity, mass,
star formation rate, specific star formation rate, UV attenuation, and
metallicity. Like LBGs, they are compact systems undergoing intense
star formation.

The question of whether this is the first major episode of star
formation that these galaxies have undergone is crucial to determining
how similar they are to LBGs. $Spitzer$ observations of LBGs in the
rest-frame near-IR have shown that they do not contain a previously
hidden population of older, low-mass stars \citep{ba04}. The
UV-optical colors of many of the compact UVLGs suggest that this may
be the first major episode of star formation these galaxies have
experienced (see Figure~\ref{sfrvifig}). Near-infrared observations
are necessary to fully trace the population of old stars. Such
observations have been carried out with $Spitzer$, and will be
reported in a forthcoming paper. The current UV-optical data, however,
suggest that the compact UVLGs are indeed excellent local analogs to
LBGs. Further study of these remarkable objects can provide crucial
information about star formation in the early universe. Toward this
end we are currently analyzing {\it Hubble Space Telescope} data (in
addition to our $Spitzer$ data) in order to study the morphologies,
dust content, and stellar populations of compact UVLGs. The results
from these studies will be presented in future papers.

Whether or not the highest surface brightness UVLGs are indeed analogs
of LBGs, they are distinct from any previously studied population of
galaxies in the local universe. They are much more luminous in the UV
(and thus have much higher star formation rates) than blue compact
dwarf galaxies (BCDGs), and have higher UV surface brightnesses (and
thus star formation rates per unit area) than local starburst
galaxies. For example, while I~Zw~18, the prototypical BCDG, has an
FUV surface brightness of
$I_{1530}=8.2\times10^7$~L$_{\odot}$~kpc$^{-2}$, which is close to the
boundary for compact UVLGs, its FUV luminosity
$L_{1530}=1.7\times10^8$~L$_{\odot}$, more than 2 orders of magnitude
below the lower limit for UVLGs \citep{gdp06}.

\begin{figure}
\epsscale{1.0}
\plotone{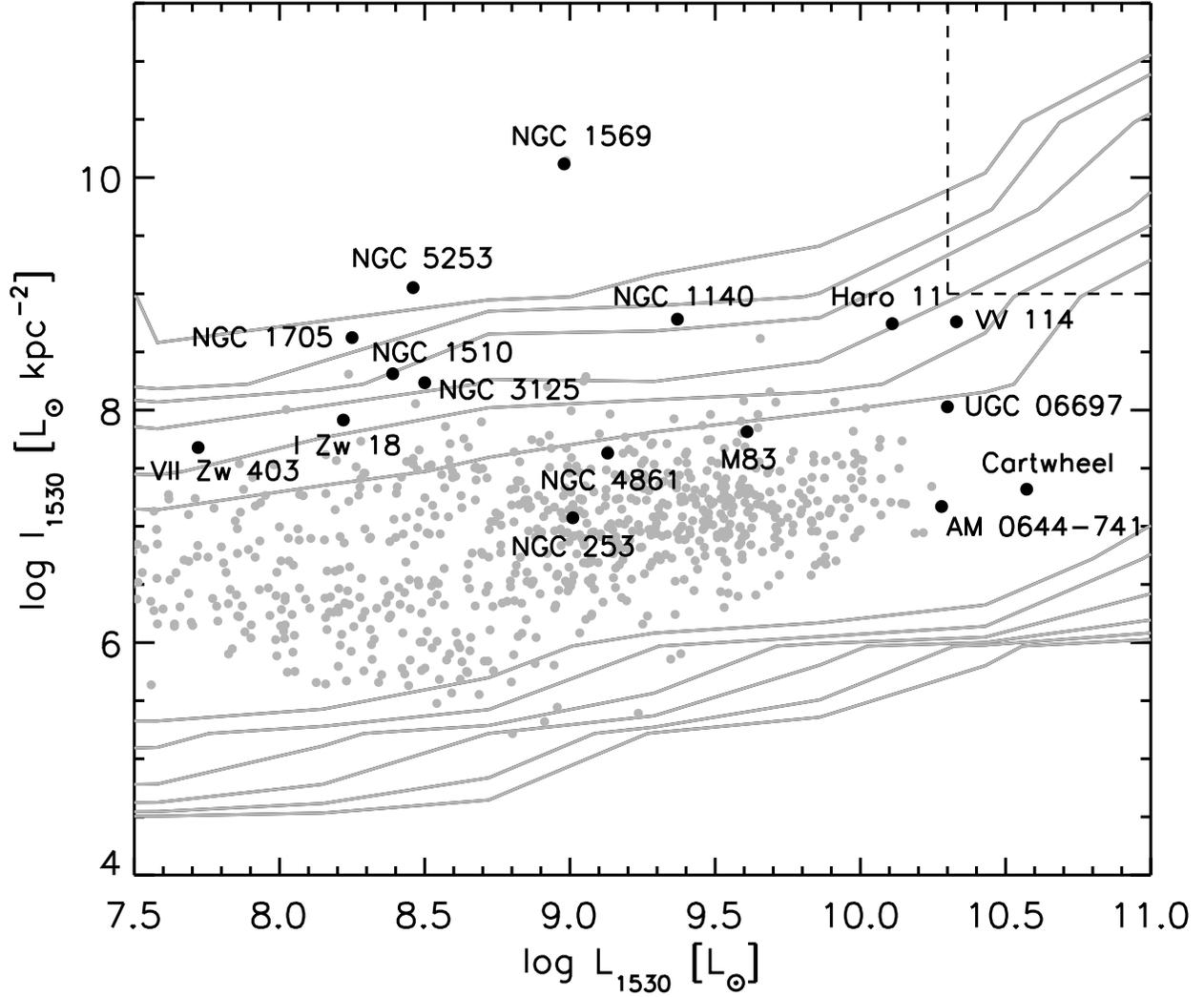}
\caption{
Same as Figure~\ref{sbfig}. UV properties for the galaxies in the
\galex\ Ultraviolet Atlas of Nearby Galaxies (Gil de Paz et al. 2006)
are shown as grey filled circles. Several local starbursts and Blue
Compact Dwarf Galaxies (BCDGs) are shown as black filled circles and
labeled. The sizes of some of the BCDGs were estimated from the GALEX
images if they were not included in the atlas. Also shown is VV~114,
the nearest example of a compact UVLG. }
\label{ngs}
\end{figure}

Figure~\ref{ngs} compares the luminosity and surface brightness of
galaxies in the \galex\ Ultraviolet Atlas of Nearby Galaxies (Gil de
Paz et al. 2006), including some nearby well-known starbursts and
BCDGs, with the normalized distribution of the entire sample.  The
range in log $L_{1530}$ for 14 starbursts from the atlas (not all
shown in the figure) is from 8.2 to 10.5 , and for 10 BCDGs in the
atlas the range is 7.3 to 10.3. The boundary for UVLGs is log
$L_{1530}$=10.3. In surface brightness the starbursts range from log
I$_{1530}=$6.2 to 9.1, while the BCDGs range from 6.6 to 8.7. The
Cartwheel galaxy, AM~0644-741, and UGC~06697 are three starburst
galaxies that are also local examples of large UVLGs. Figure~\ref{ngs}
shows that while some local galaxies have high FUV surface brightness,
and others have high FUV luminosity, almost none have the combination
of high luminosity and high surface brightness that would qualify as a
compact UVLG. The exception is VV~114, which is the nearest known
Lyman Break Galaxy analog \citep{grimes06}. Compact UVLGs, and
especially supercompact UVLGs, clearly have extreme properties when
compared to local galaxy populations.

\section{Conclusions}	

We have used \galex\ and the SDSS to identify and study the most UV
luminous galaxies in the local universe. Our main results are as
follows:

\begin{enumerate}

\item{
The most UV luminous galaxies in the local universe comprise a diverse
group, with properties that are well correlated with UV surface
brightness. Although there is not a sharp transition, we can use a
surface brightness boundary of $I_{1530}<10^8$~L$_{\odot}$~kpc$^{-2}$
to divide the UVLG sample into two groups: large and compact UVLGs.}

\item{
The large UVLGs are massive, metal-rich disk galaxies with UV
surface brightnesses only slightly larger than typical star forming
galaxies. They are similar to normal galaxies in most respects, but
are very luminous primarily because of their large size.}

\item{ 
The compact UVLGs are low-mass, relatively metal-poor systems which
often have a disturbed or interacting morphology. The high UV surface
brightness and high specific star formation rate in these compact
UVLGs indicates that intense star formation is ongoing.}

\item{
It is possible to isolate a sample of local LBG-analogs with UV surface
brightness criterion of $I_{1530}>10^9$~L$_{\odot}$~kpc$^{-2}$. The
galaxies in the resulting sample have many properties in common with
LBGs, including luminosity, mass, star formation rate, specific star
formation rate, extinction, and metallicity.}

\item{
Compact UVLGs stand out from the trends established by the full
DR3/GR1 sample in that they have much smaller sizes and higher surface
brightness than would be predicted for galaxies of their mass or
luminosity. They have metallicities that are generally lower by a
factor of two to three compared to normal galaxies of the same mass.
These properties suggest that they are a distinct population of
objects, perhaps at a different phase of evolution than the bulk of
the galaxies in the local universe.}

\item{
The high UV luminosity and implied high star formation rate of compact
UVLGs distinguishes them from any previously studied local galaxy
population, including local UV-bright starbursts and blue compact
dwarf galaxies.}

\end{enumerate} 

\acknowledgments

We thank the referee, Michael Strauss, for providing very helpful
comments that greatly improved the paper. \galex\ is a NASA Small
Explorer launched in April 2003. We gratefully acknowledge NASA's
support for construction, operation, and scientific analysis for the
\galex\ mission. Funding for the creation and distribution of the SDSS
Archive has been provided by the Alfred P. Sloan Foundation, the
Participating Institutions, the National Aeronautics and Space
Administration, the National Science Foundation, the U.S. Department
of Energy, the Japanese Monbukagakusho, and the Max Planck Society.



\begin{deluxetable}{lcccccccccccc}
\rotate
\tabletypesize{\scriptsize}
\tablecaption{UVLG Sample \label{samptab}}
\tablewidth{0pt}
\tablehead{
\colhead{SDSS ObjID} & \colhead{RA} & \colhead{Dec} & \colhead{Redshift} & \colhead{log L$_{1530}$} & \colhead{r$_{50,u}$} & \colhead{log I$_{1530}$} & \colhead{FUV} & \colhead{NUV} & \colhead{A$_{1530}$} & \colhead{log M$_*$} & \colhead{log SFR} & \colhead{12+log(O/H)} \\
\colhead{} & \colhead{(deg)} & \colhead{(deg)} & \colhead{} & \colhead{(L$_{\odot}$)} & \colhead{(kpc)}& \colhead{(L$_{\odot}$ kpc$^{-2}$)}& \colhead{(mag)} & \colhead{(mag)} & \colhead{(mag)}  & \colhead{(M$_{\odot}$)} & \colhead{(M$_{\odot}$ yr$^{-1}$)} & \colhead{} \\
\colhead{(1)} & \colhead{(2)} & \colhead{(3)} & \colhead{(4)} & \colhead{(5)} & \colhead{(6)}& \colhead{(7)}& \colhead{(8)} & \colhead{(9)} & \colhead{(10)}  & \colhead{(11)} & \colhead{(12)} & \colhead{(13)} }
\startdata
587731187814629504 &   0.20128 &   1.15675 & 0.250 & 10.34 &  3.13 &  8.55 & 20.51$\pm$0.25 & 20.16$\pm$0.15 &  0.80 & 10.08 &  1.03 &  8.89 \\    
588015508196294786 &   1.46632 &  -0.73521 & 0.134 & 10.34 &  9.62 &  7.58 & 19.02$\pm$0.28 & 20.09$\pm$0.26 &  3.25 & 10.76 &  0.76 & \nodata \\  
588015507659554916 &   1.77391 &  -1.07124 & 0.288 & 10.51 &  5.14 &  8.29 & 20.44$\pm$0.25 & 20.34$\pm$0.17 &  1.51 & 10.67 &  1.44 &  8.89 \\    
587731185131192547 &   2.40203 &  -0.84206 & 0.161 & 10.42 &  7.69 &  7.85 & 19.26$\pm$0.15 & 18.87$\pm$0.08 &  2.26 & 10.56 &  0.60 &  9.02 \\    
587727225154306171 &   2.45330 & -11.03518 & 0.211 & 10.35 &  4.13 &  8.32 & 20.08$\pm$0.22 & 19.55$\pm$0.10 &  1.42 &  9.98 &  1.07 &  8.67 \\    
588015508196753479 &   2.54157 &  -0.76765 & 0.243 & 10.45 &  1.38 &  9.37 & 20.19$\pm$0.23 & 19.86$\pm$0.15 &  1.00 & 10.43 &  1.18 &  8.74 \\    
588015510344302646 &   2.78681 &   0.84544 & 0.108 & 10.54 &  3.24 &  8.72 & 18.02$\pm$0.08 & 17.87$\pm$0.04 &  0.37 &  9.76 &  1.71 &  8.71 \\    
588015509270626537 &   2.86065 &   0.06754 & 0.211 & 10.36 &  6.55 &  7.93 & 20.07$\pm$0.22 & 19.86$\pm$0.11 &  2.96 & 10.62 &  1.15 &  9.02 \\    
587727225154764940 &   3.43824 & -10.93658 & 0.246 & 10.57 & 12.52 &  7.58 & 19.90$\pm$0.08 & 19.27$\pm$0.03 &  2.35 & 10.99 &  1.20 &  9.06 \\    
587727225154961583 &   3.99757 & -11.02590 & 0.267 & 10.40 & 23.68 &  6.85 & 20.53$\pm$0.07 & 19.99$\pm$0.03 &  2.21 & 11.62 &  0.95 & \nodata \\  
587731185132109826 &   4.36789 &  -0.94027 & 0.094 & 10.32 &  2.70 &  8.66 & 18.25$\pm$0.09 & 17.92$\pm$0.05 &  1.73 & 10.74 &  1.25 &  8.96 \\    
587731185132241048 &   4.78147 &  -0.87711 & 0.187 & 10.42 &  4.13 &  8.39 & 19.63$\pm$0.17 & 19.35$\pm$0.10 &  1.37 & 10.28 &  1.35 &  8.89 \\    
587727227304476771 &   8.60397 &  -9.19877 & 0.164 & 10.47 &  8.98 &  7.77 & 19.17$\pm$0.04 & 19.57$\pm$0.04 &  1.70 & 10.76 &  0.45 & \nodata \\  
587727180600705169 &   8.71946 &  -9.02820 & 0.144 & 10.38 & 10.40 &  7.55 & 19.09$\pm$0.14 & 18.83$\pm$0.07 &  2.27 & 10.27 &  0.66 &  8.96 \\    
587727177916416108 &   8.84842 & -11.12945 & 0.123 & 10.33 &  9.98 &  7.53 & 18.83$\pm$0.03 & 20.19$\pm$0.03 &  2.85 & 10.44 & -0.13 & \nodata \\  
587724198812057678 &   9.06934 &  15.04284 & 0.119 & 10.63 &  3.99 &  8.63 & 18.02$\pm$0.03 & \nodata &  1.91 & 10.61 &  0.21 &  9.11 \\    
587727225157386334 &   9.58818 & -10.88660 & 0.184 & 10.52 &  9.26 &  7.79 & 19.34$\pm$0.15 & 19.13$\pm$0.08 &  2.18 & 10.77 &  1.06 &  8.94 \\    
587724199349387411 &  10.22635 &  15.56938 & 0.283 & 10.43 &  1.72 &  9.16 & 20.62$\pm$0.08 & 21.03$\pm$0.05 &  0.13 &  9.18 &  0.41 &  8.23 \\    
587724199349518558 &  10.52352 &  15.36599 & 0.259 & 10.32 & 13.81 &  7.24 & 20.66$\pm$0.10 & 20.28$\pm$0.04 &  9.20 & 11.27 &  0.91 & \nodata \\  
588015509274165442 &  10.94763 &   0.05960 & 0.213 & 10.83 & 13.08 &  7.80 & 18.92$\pm$0.11 & 18.94$\pm$0.08 &  0.63 & 11.24 &  0.84 & \nodata \\  
587724199349780648 &  11.19722 &  15.48662 & 0.227 & 10.56 &  2.30 &  9.04 & 19.75$\pm$0.05 & 19.52$\pm$0.03 & \nodata & 10.71 &  1.39 & \nodata \\
588015510348169371 &  11.61157 &   0.86546 & 0.166 & 10.42 &  2.86 &  8.71 & 19.34$\pm$0.14 & 18.74$\pm$0.06 &  1.31 & 10.41 &  1.13 &  8.60 \\    
587724233712140399 &  11.69467 &  15.72774 & 0.181 & 10.43 &  9.28 &  7.70 & 19.52$\pm$0.05 & 19.16$\pm$0.02 &  3.41 & 11.10 &  1.06 &  9.27 \\    
587724232101724283 &  12.19103 &  14.44148 & 0.278 & 10.40 &  4.04 &  8.39 & 20.64$\pm$0.31 & 20.46$\pm$0.21 &  1.97 & 11.00 &  1.92 & \nodata \\  
588015508738539596 &  13.86443 &  -0.36355 & 0.167 & 10.49 &  0.80 &  9.89 & 19.19$\pm$0.14 & 18.72$\pm$0.07 & \nodata & 10.07 &  1.18 & \nodata \\
587727179529322564 &  14.14723 &  -9.82308 & 0.189 & 10.58 &  4.74 &  8.43 & 19.23$\pm$0.04 & \nodata &  0.00 & 11.08 &  0.57 &  9.01 \\           
587731187283853386 &  14.19609 &   0.64680 & 0.218 & 10.43 &  2.38 &  8.88 & 19.96$\pm$0.20 & 19.30$\pm$0.12 &  1.38 & 10.02 &  0.92 &  8.65 \\    
587724234250256545 &  14.69203 &  16.03578 & 0.243 & 10.36 &  4.18 &  8.32 & 20.41$\pm$0.28 & 19.76$\pm$0.16 &  1.47 & 10.17 &  0.78 &  8.72 \\    
587724199888289961 &  14.98405 &  15.64598 & 0.144 & 11.16 & 11.24 &  8.26 & 17.14$\pm$0.02 & 19.11$\pm$0.02 &  1.99 & 10.94 &  0.53 & \nodata \\  
587724197204132105 &  15.36074 &  13.54601 & 0.223 & 10.51 & 11.28 &  7.61 & 19.82$\pm$0.20 & 19.72$\pm$0.15 & 10.44 & 11.10 &  0.69 & \nodata \\  
587724198814810262 &  15.61241 &  14.91062 & 0.086 & 10.33 &  1.68 &  9.08 & 18.00$\pm$0.03 & \nodata &  0.66 &  9.93 & -0.29 &  8.89 \\           
587731511530815637 &  16.03357 &  -0.84219 & 0.207 & 10.38 &  4.08 &  8.36 & 19.96$\pm$0.23 & 19.29$\pm$0.09 &  0.80 & 10.09 &  1.44 &  8.78 \\    
588015509276590315 &  16.56369 &   0.12798 & 0.220 & 10.41 &  8.69 &  7.73 & 20.04$\pm$0.22 & 19.94$\pm$0.12 &  1.83 & 10.86 &  0.78 &  9.11 \\    
587731514215497820 &  16.70578 &   1.05622 & 0.253 & 10.95 &  5.02 &  8.75 & 19.04$\pm$0.14 & 19.69$\pm$0.11 &  1.43 & 11.70 &  2.41 & \nodata \\  
587731514215497878 &  16.80949 &   1.15869 & 0.159 & 10.34 &  6.82 &  7.87 & 19.43$\pm$0.17 & 19.55$\pm$0.10 &  2.01 & 10.78 &  0.91 &  9.14 \\    
587724197204983886 &  17.31450 &  13.43438 & 0.184 & 10.55 & 12.58 &  7.55 & 19.25$\pm$0.16 & 19.02$\pm$0.09 & \nodata & 11.48 &  1.63 & \nodata \\
587724233178808483 &  20.14169 &  14.77605 & 0.130 & 10.48 & 11.91 &  7.53 & 18.61$\pm$0.03 & \nodata &  3.26 & 11.07 &  0.53 & \nodata \\         
587724233179332766 &  21.25810 &  14.63618 & 0.153 & 10.37 &  8.52 &  7.71 & 19.26$\pm$0.04 & 18.81$\pm$0.02 &  2.63 & 11.12 &  1.23 & \nodata \\  
587727180069404773 &  21.54104 &  -9.12973 & 0.263 & 10.38 & 30.63 &  6.61 & 20.55$\pm$0.07 & 20.08$\pm$0.05 &  2.23 & 11.70 &  0.99 & \nodata \\  
587727178460496000 &  25.59877 & -10.18952 & 0.128 & 10.41 &  3.47 &  8.53 & 18.72$\pm$0.03 & 18.41$\pm$0.01 &  0.56 &  9.56 &  0.55 &  8.67 \\    
587724233181429914 &  26.26956 &  14.25279 & 0.172 & 10.36 &  9.84 &  7.58 & 19.57$\pm$0.05 & 19.36$\pm$0.02 &  0.00 & 11.14 & -0.20 & \nodata \\  
587727229449338983 &  26.80144 & -10.21155 & 0.198 & 10.35 & 14.62 &  7.22 & 19.92$\pm$0.05 & 19.31$\pm$0.02 &  0.00 & 11.06 &  0.86 & \nodata \\  
587724232644821098 &  26.81554 &  13.91049 & 0.193 & 10.47 &  3.21 &  8.66 & 19.58$\pm$0.05 & 19.03$\pm$0.02 &  0.48 & 10.08 &  1.44 &  8.70 \\    
587724197746180186 &  27.61831 &  13.14958 & 0.147 & 10.62 &  1.66 &  9.38 & 18.55$\pm$0.03 & 18.26$\pm$0.02 &  1.26 & 10.42 &  0.91 &  8.87 \\    
587724232108409001 &  27.85828 &  13.41965 & 0.243 & 10.52 &  1.53 &  9.35 & 20.00$\pm$0.06 & 19.53$\pm$0.03 &  1.19 & 10.52 &  1.29 &  8.98 \\    
587724197746376839 &  27.95274 &  13.19747 & 0.191 & 10.60 &  7.58 &  8.04 & 19.21$\pm$0.05 & \nodata &  2.21 & 11.32 & -0.01 & \nodata \\         
587724233720004940 &  30.32127 &  14.33144 & 0.285 & 10.34 & 54.49 &  6.07 & 20.86$\pm$0.09 & 19.91$\pm$0.03 &  3.55 & 11.10 & -0.07 &  9.10 \\    
587724198822215805 &  32.98133 &  13.47305 & 0.149 & 10.33 & 14.74 &  7.20 & 19.30$\pm$0.05 & 18.83$\pm$0.02 &  1.79 & 11.23 &  0.90 & \nodata \\  
587724232647508116 &  33.09660 &  13.20805 & 0.134 & 10.34 & 13.95 &  7.25 & 19.02$\pm$0.04 & 18.59$\pm$0.02 &  3.75 & 11.21 &  0.99 & \nodata \\  
587724198285541435 &  33.45223 &  12.99766 & 0.219 & 10.58 &  0.85 &  9.92 & 19.59$\pm$0.06 & 18.72$\pm$0.02 &  2.48 & 10.68 &  1.35 & \nodata \\  
587731511538745437 &  34.13192 &  -1.02454 & 0.167 & 10.39 &  3.43 &  8.52 & 19.41$\pm$0.16 & 18.88$\pm$0.07 &  1.44 & 10.34 &  1.51 &  8.85 \\    
587727179540463751 &  39.85733 &  -8.14107 & 0.184 & 10.99 & 13.21 &  7.95 & 18.16$\pm$0.03 & 18.71$\pm$0.01 &  6.02 & 11.40 &  0.06 & \nodata \\  
587727179540594732 &  40.14347 &  -8.15686 & 0.177 & 10.31 &  5.77 &  7.99 & 19.74$\pm$0.05 & 19.27$\pm$0.02 &  1.36 & 10.51 &  0.76 &  9.00 \\    
587724241226498139 &  41.37314 &  -8.27718 & 0.195 & 10.41 &  1.66 &  9.17 & 19.74$\pm$0.05 & 19.08$\pm$0.03 &  1.39 & 10.36 &  1.36 &  8.83 \\    
587724240153018558 &  42.18780 &  -8.96868 & 0.238 & 10.73 & 20.17 &  7.32 & 19.42$\pm$0.04 & 19.28$\pm$0.03 &  2.33 & 10.99 &  0.98 &  9.00 \\    
587724241226891413 &  42.38908 &  -8.09707 & 0.270 & 11.11 & 24.91 &  7.52 & 18.77$\pm$0.03 & 18.63$\pm$0.02 &  2.03 & 10.73 &  1.78 &  8.61 \\    
587731511543464076 &  44.95400 &  -0.96678 & 0.293 & 10.59 &  5.14 &  8.37 & 20.30$\pm$0.23 & 19.47$\pm$0.11 &  1.95 & 10.96 &  1.93 &  8.87 \\    
588015509825978540 &  45.13415 &   0.51732 & 0.170 & 10.38 &  3.36 &  8.53 & 19.48$\pm$0.14 & 19.76$\pm$0.11 &  2.61 & 10.84 &  1.00 & \nodata \\  
587727180616695902 &  45.53338 &  -6.63504 & 0.180 & 10.42 & 12.46 &  7.43 & 19.52$\pm$0.20 & 19.18$\pm$0.10 &  1.62 & 10.94 &  0.85 &  8.87 \\    
587727179543740592 &  47.44834 &  -7.33566 & 0.196 & 10.43 &  4.96 &  8.24 & 19.71$\pm$0.06 & 19.58$\pm$0.03 &  2.26 & 10.58 &  0.95 &  9.06 \\    
587724241766187109 &  47.84834 &  -7.00088 & 0.226 & 10.44 & 11.45 &  7.52 & 20.02$\pm$0.06 & 19.76$\pm$0.03 &  2.14 & 11.33 &  0.83 &  8.98 \\    
587724241229643897 &  48.61778 &  -7.42162 & 0.208 & 10.32 &  2.02 &  8.91 & 20.12$\pm$0.08 & 20.09$\pm$0.04 &  0.78 & 10.46 &  0.29 & \nodata \\  
587724240156360834 &  49.81060 &  -8.06979 & 0.125 & 10.50 &  3.21 &  8.69 & 18.45$\pm$0.12 & 18.25$\pm$0.07 &  1.79 & 10.29 &  1.00 &  8.88 \\    
587731511545954498 &  50.61609 &  -1.03658 & 0.194 & 10.31 & 14.16 &  7.21 & 19.98$\pm$0.28 & 20.19$\pm$0.30 &  2.94 & 10.94 &  0.51 & \nodata \\  
587731514230571148 &  51.21346 &   1.11823 & 0.150 & 10.31 & 12.84 &  7.29 & 19.35$\pm$0.20 & 19.24$\pm$0.20 &  1.10 & 11.03 &  0.31 & \nodata \\  
588015510365602004 &  51.34779 &   0.96633 & 0.212 & 10.33 & 12.41 &  7.34 & 20.15$\pm$0.29 & 20.30$\pm$0.26 &  0.53 & 11.21 &  0.80 &  9.31 \\    
587731512083415213 &  51.98476 &  -0.47176 & 0.162 & 10.54 &  4.53 &  8.43 & 18.97$\pm$0.16 & 18.62$\pm$0.09 &  0.90 & 10.29 &  1.64 &  8.83 \\    
587731512620351653 &  52.09571 &  -0.13294 & 0.205 & 10.31 &  3.32 &  8.47 & 20.12$\pm$0.30 & 19.70$\pm$0.18 &  0.81 &  9.95 &  1.28 &  8.85 \\    
587731514231029883 &  52.19166 &   1.19744 & 0.142 & 10.39 &  1.74 &  9.11 & 19.04$\pm$0.17 & 19.05$\pm$0.12 &  0.48 &  9.74 &  1.19 &  8.78 \\    
587724242305286402 &  52.93439 &  -5.93593 & 0.247 & 10.52 &  2.76 &  8.84 & 20.03$\pm$0.06 & 19.60$\pm$0.04 &  0.33 &  9.79 &  1.31 &  8.61 \\    
587731513694617768 &  53.22892 &   0.64229 & 0.227 & 10.54 & 12.33 &  7.56 & 19.78$\pm$0.24 & 19.42$\pm$0.17 &  1.51 & 11.14 &  0.89 &  9.02 \\    
588015509292712195 &  53.34551 &   0.09876 & 0.178 & 10.49 &  8.30 &  7.85 & 19.32$\pm$0.17 & \nodata &  4.56 & 11.03 &  0.57 & \nodata \\         
587731514231750803 &  53.87266 &   1.24974 & 0.171 & 10.48 &  6.96 &  8.00 & 19.25$\pm$0.21 & 19.93$\pm$0.24 &  1.49 & 10.63 &  0.68 &  9.14 \\    
587731514232144016 &  54.77985 &   1.16598 & 0.182 & 10.36 &  6.40 &  7.95 & 19.71$\pm$0.25 & 20.03$\pm$0.31 &  2.33 & 10.89 &  0.68 &  9.03 \\    
587731514232144028 &  54.80193 &   1.13992 & 0.261 & 10.62 &  3.60 &  8.71 & 19.94$\pm$0.27 & 19.64$\pm$0.18 &  2.05 & 10.90 &  1.15 &  9.07 \\    
588015510367240313 &  55.09066 &   0.92732 & 0.253 & 10.36 &  5.70 &  8.05 & 20.52$\pm$0.30 & 19.75$\pm$0.20 &  1.43 & 10.59 &  1.07 &  9.03 \\    
587731513158991989 &  56.09607 &   0.39638 & 0.219 & 10.32 &  4.04 &  8.31 & 20.25$\pm$0.30 & 19.78$\pm$0.14 &  1.53 & 10.33 &  1.60 &  8.82 \\    
587724242307121309 &  57.03210 &  -5.42077 & 0.164 & 10.38 &  4.91 &  8.20 & 19.38$\pm$0.19 & 19.19$\pm$0.12 &  0.62 & 10.04 &  1.17 &  8.58 \\    
587724240697622665 &  59.81242 &  -6.30119 & 0.181 & 10.51 &  8.05 &  7.90 & 19.30$\pm$0.19 & 19.21$\pm$0.15 &  3.03 & 11.07 &  1.41 &  8.98 \\    
587727180086378659 &  60.53695 &  -5.11169 & 0.139 & 10.45 &  1.20 &  9.49 & 18.85$\pm$0.15 & 18.88$\pm$0.09 &  0.23 &  9.39 &  1.22 &  8.53 \\    
588007005229875657 & 115.76499 &  37.73554 & 0.202 & 10.44 &  5.97 &  8.09 & 19.74$\pm$0.05 & 19.31$\pm$0.02 &  2.13 & 11.44 &  1.05 &  9.11 \\    
587735236344414396 & 120.41840 &  23.20090 & 0.272 & 10.33 &  3.85 &  8.36 & 20.76$\pm$0.30 & 19.65$\pm$0.10 &  2.69 & 10.78 &  1.31 &  9.06 \\    
587728668268167253 & 122.18443 &  39.81455 & 0.091 & 10.45 &  0.50 & 10.25 & 17.85$\pm$0.08 & 17.54$\pm$0.04 &  1.04 & 10.33 &  1.02 & \nodata \\  
587725981224599567 & 123.84753 &  50.07076 & 0.164 & 10.50 &  1.87 &  9.16 & 19.11$\pm$0.15 & 18.61$\pm$0.07 &  1.51 & 10.30 &  2.10 &  8.99 \\    
587725774535000285 & 124.04864 &  49.38725 & 0.179 & 10.47 & 12.72 &  7.46 & 19.37$\pm$0.17 & 19.17$\pm$0.09 &  3.12 & 11.30 &  1.23 & \nodata \\  
587725470130110508 & 124.34376 &  46.74989 & 0.280 & 10.83 &  3.24 &  9.01 & 19.57$\pm$0.05 & 19.13$\pm$0.02 &  0.48 &  9.98 &  1.53 &  8.68 \\    
588007004697854025 & 124.67503 &  46.58495 & 0.218 & 10.43 &  2.24 &  8.93 & 19.97$\pm$0.05 & 19.31$\pm$0.02 &  3.02 & 11.21 &  2.21 &  9.06 \\    
587728668806611190 & 124.74548 &  42.90008 & 0.154 & 10.43 & 13.09 &  7.40 & 19.13$\pm$0.15 & 18.87$\pm$0.08 &  1.73 & 11.25 &  0.77 & \nodata \\  
587728931875651786 & 124.86896 &  41.29362 & 0.179 & 10.41 &  6.86 &  7.94 & 19.55$\pm$0.18 & 19.88$\pm$0.12 &  1.96 & 10.82 &  1.42 &  9.11 \\    
587728664505483532 & 124.88740 &  42.60294 & 0.232 & 10.53 &  2.86 &  8.82 & 19.86$\pm$0.21 & 20.28$\pm$0.15 &  1.20 & 10.98 &  0.50 & \nodata \\  
588007004160983209 & 125.22800 &  46.21449 & 0.229 & 10.41 &  2.96 &  8.67 & 20.13$\pm$0.06 & 19.63$\pm$0.03 &  1.16 &  9.79 &  1.13 &  8.66 \\    
587731872314884117 & 125.40520 &  37.17965 & 0.284 & 10.80 &  1.84 &  9.47 & 19.70$\pm$0.12 & 19.27$\pm$0.07 &  0.58 &  9.99 &  1.54 &  8.89 \\    
587728664506073210 & 126.05471 &  43.62250 & 0.118 & 10.61 &  2.64 &  8.97 & 18.05$\pm$0.09 & 17.90$\pm$0.05 & \nodata & 10.36 &  1.40 & \nodata \\
587728931876176092 & 126.06725 &  42.16934 & 0.223 & 10.34 & 10.02 &  7.54 & 20.24$\pm$0.24 & \nodata &  2.11 & 11.13 & -2.05 & \nodata \\         
587731885736788122 & 126.13889 &  38.00366 & 0.103 & 10.33 &  4.37 &  8.25 & 18.43$\pm$0.07 & 17.84$\pm$0.03 &  1.22 & 10.93 &  1.71 &  9.15 \\    
587728664506400946 & 126.71116 &  44.14319 & 0.132 & 10.35 & 11.51 &  7.43 & 18.97$\pm$0.13 & 18.31$\pm$0.07 &  1.53 & 11.05 &  0.82 & \nodata \\  
587731886275166445 & 128.88380 &  40.71084 & 0.084 & 10.31 &  4.06 &  8.30 & 18.01$\pm$0.08 & 17.94$\pm$0.05 &  0.44 &  9.67 &  0.52 &  8.61 \\    
588010136800395436 & 129.38982 &  47.96454 & 0.215 & 10.56 &  2.96 &  8.82 & 19.59$\pm$0.15 & 18.93$\pm$0.06 &  0.30 &  9.91 &  1.41 &  8.31 \\    
588010135726456918 & 129.61571 &  47.17575 & 0.097 & 10.38 &  2.07 &  8.95 & 18.17$\pm$0.01 & 17.73$\pm$0.01 &  1.35 & 10.72 &  0.80 &  8.88 \\    
588009365859074276 & 130.71088 &  46.97751 & 0.229 & 10.33 & 11.08 &  7.44 & 20.33$\pm$0.04 & 19.66$\pm$0.02 &  2.79 & 11.15 &  1.09 & \nodata \\  
587728931341336751 & 130.91106 &  45.25332 & 0.193 & 10.42 &  4.74 &  8.27 & 19.68$\pm$0.16 & 19.46$\pm$0.08 &  2.39 & 10.65 &  1.43 &  9.10 \\    
587725470670192770 & 131.50928 &  52.53307 & 0.053 & 10.31 &  3.95 &  8.32 & 16.95$\pm$0.01 & 16.50$\pm$0.00 &  2.24 & 11.15 &  1.06 &  9.20 \\    
587728930804859108 & 132.25192 &  45.56725 & 0.220 & 10.58 & 10.79 &  7.72 & 19.61$\pm$0.16 & 19.34$\pm$0.08 &  2.85 & 11.03 &  0.88 &  9.08 \\    
587725552275226831 & 133.40443 &  57.07639 & 0.299 & 11.33 & 32.03 &  7.52 & 18.48$\pm$0.03 & \nodata & \nodata & 11.53 & -0.17 & \nodata \\       
587732049480056879 & 136.07295 &  45.62538 & 0.195 & 10.47 &  9.68 &  7.70 & 19.58$\pm$0.20 & 19.34$\pm$0.10 &  2.70 & 10.85 &  0.99 &  9.11 \\    
587725469598023726 & 136.63254 &  54.47703 & 0.205 & 10.41 &  2.58 &  8.79 & 19.86$\pm$0.17 & 19.23$\pm$0.08 &  1.00 & 10.15 &  1.61 &  8.85 \\    
587732702855627062 & 136.67363 &   5.37485 & 0.226 & 10.33 &  9.72 &  7.56 & 20.30$\pm$0.26 & 20.07$\pm$0.17 &  2.80 & 11.29 &  0.99 & \nodata \\  
587732701782016286 & 137.08205 &   4.58679 & 0.124 & 10.35 &  9.05 &  7.64 & 18.82$\pm$0.14 & 18.24$\pm$0.06 &  2.43 & 10.74 &  0.90 &  9.06 \\    
588007004703096884 & 137.59975 &  55.39899 & 0.171 & 10.33 &  2.24 &  8.83 & 19.62$\pm$0.04 & 19.04$\pm$0.02 &  2.21 & 11.00 &  1.37 & \nodata \\  
587727942954844494 & 137.80559 &   1.23246 & 0.252 & 10.32 & 32.82 &  6.49 & 20.60$\pm$0.07 & 20.29$\pm$0.03 &  4.55 & 11.31 &  0.53 & \nodata \\  
587732054319890553 & 139.35410 &  42.75827 & 0.237 & 10.31 &  7.26 &  7.79 & 20.45$\pm$0.22 & 19.81$\pm$0.11 &  2.50 & 11.09 &  1.27 & \nodata \\  
587725817477988424 & 139.44617 &  59.29116 & 0.173 & 10.38 &  2.43 &  8.81 & 19.54$\pm$0.17 & 18.66$\pm$0.07 &  1.78 & 10.56 &  1.27 &  8.91 \\    
587731913110257902 & 139.95036 &  48.64304 & 0.237 & 10.32 & 14.44 &  7.20 & 20.43$\pm$0.25 & 19.88$\pm$0.11 &  1.76 & 11.01 &  0.91 &  9.04 \\    
587725551740321978 & 140.07605 &  59.86119 & 0.195 & 10.39 & 10.64 &  7.54 & 19.78$\pm$0.19 & 19.71$\pm$0.13 &  2.34 & 10.90 &  0.77 &  9.05 \\    
587731521208582257 & 140.49748 &  45.15344 & 0.235 & 10.82 &  1.53 &  9.65 & 19.17$\pm$0.13 & 18.80$\pm$0.06 &  2.22 & 10.54 &  1.99 & \nodata \\  
587725551203451103 & 140.60236 &  59.61079 & 0.277 & 10.40 & 16.56 &  7.16 & 20.63$\pm$0.27 & 20.18$\pm$0.17 &  2.30 & 11.47 &  0.57 & \nodata \\  
588007004167012515 & 140.65002 &  56.27194 & 0.191 & 10.32 &  6.25 &  7.93 & 19.91$\pm$0.19 & 19.23$\pm$0.08 &  3.15 & 11.10 &  1.28 &  9.11 \\    
587731913110520003 & 140.65392 &  48.95396 & 0.185 & 10.42 & 16.01 &  7.21 & 19.60$\pm$0.16 & 19.10$\pm$0.07 &  0.84 & 11.09 &  0.58 & \nodata \\  
588010136804786207 & 140.90192 &  54.81091 & 0.222 & 10.56 &  1.14 &  9.65 & 19.68$\pm$0.17 & 19.45$\pm$0.10 &  1.36 &  9.97 &  0.85 &  8.86 \\    
587725470673535009 & 141.13921 &  57.73800 & 0.227 & 10.69 &  9.29 &  7.96 & 19.42$\pm$0.16 & \nodata &  2.28 & 11.18 &  0.75 &  8.98 \\           
588013384341913605 & 141.50169 &  44.46005 & 0.181 & 10.73 &  1.03 &  9.91 & 18.76$\pm$0.10 & 18.84$\pm$0.06 &  0.46 &  9.20 &  0.72 &  8.46 \\    
587731521746043150 & 141.68233 &  46.17589 & 0.154 & 10.31 & 12.61 &  7.31 & 19.41$\pm$0.15 & 19.19$\pm$0.08 &  1.45 & 10.55 &  0.64 & \nodata \\  
588007006315544766 & 142.18282 &  59.00600 & 0.226 & 10.62 & 27.68 &  6.94 & 19.57$\pm$0.04 & 19.34$\pm$0.02 &  0.34 & 10.54 &  2.46 &  8.77 \\    
587729385531310202 & 142.37518 &  50.58275 & 0.223 & 10.46 & 11.44 &  7.54 & 19.94$\pm$0.22 & 19.62$\pm$0.10 &  2.87 & 10.98 &  0.89 &  9.10 \\    
587731680122110146 & 142.46309 &  46.42466 & 0.188 & 10.33 &  8.15 &  7.71 & 19.85$\pm$0.18 & \nodata &  0.90 & 10.26 &  0.98 &  8.86 \\           
587725816405098641 & 143.44133 &  59.84856 & 0.231 & 10.37 &  1.98 &  8.98 & 20.25$\pm$0.23 & 19.86$\pm$0.11 &  0.41 &  9.70 &  1.00 &  8.70 \\    
587732484347723780 & 143.78641 &  42.78500 & 0.234 & 10.59 &  2.76 &  8.91 & 19.74$\pm$0.17 & 19.36$\pm$0.09 &  1.62 & 10.21 &  1.36 &  8.81 \\    
587728932420255860 & 143.99200 &  53.45718 & 0.199 & 10.38 &  8.38 &  7.74 & 19.86$\pm$0.17 & 19.45$\pm$0.08 &  1.55 & 10.86 &  1.05 &  9.11 \\    
588009365863989360 & 143.99982 &  54.36830 & 0.262 & 10.69 &  5.77 &  8.37 & 19.76$\pm$0.17 & 19.17$\pm$0.07 &  1.71 & 10.88 &  1.64 & \nodata \\  
588009367475847371 & 147.05156 &  56.99398 & 0.214 & 10.31 & 14.44 &  7.19 & 20.20$\pm$0.21 & 20.26$\pm$0.12 &  2.64 & 11.00 &  1.03 & \nodata \\  
587725816943083550 & 147.10237 &  61.66580 & 0.173 & 10.43 &  1.04 &  9.60 & 19.39$\pm$0.04 & 19.28$\pm$0.02 & \nodata &  9.88 &  0.92 & \nodata \\
587725816406081719 & 147.16391 &  61.22299 & 0.185 & 10.91 &  8.46 &  8.26 & 18.36$\pm$0.03 & \nodata &  1.95 & 10.82 &  0.24 & \nodata \\         
587731868014084272 & 147.40846 &  47.39235 & 0.249 & 10.36 &  6.32 &  7.96 & 20.47$\pm$0.23 & 20.51$\pm$0.17 &  2.20 & 10.61 &  1.10 &  9.06 \\    
587731869088219192 & 147.90614 &  48.66147 & 0.135 & 10.39 &  1.89 &  9.04 & 18.91$\pm$0.11 & 18.99$\pm$0.07 &  0.56 &  9.46 &  0.49 &  8.25 \\    
588010135732814044 & 147.92696 &  56.43008 & 0.283 & 10.38 &  6.04 &  8.02 & 20.72$\pm$0.27 & 20.31$\pm$0.12 &  1.48 & 10.68 &  1.07 &  9.11 \\    
588009367476174965 & 148.14238 &  57.40535 & 0.257 & 10.49 &  3.90 &  8.51 & 20.22$\pm$0.21 & 19.71$\pm$0.09 &  1.74 & 10.78 &  2.35 &  9.00 \\    
587732049484513320 & 148.64407 &  51.58557 & 0.130 & 10.78 &  1.55 &  9.60 & 17.84$\pm$0.07 & \nodata &  0.89 &  9.56 &  1.35 &  8.63 \\           
587735241719021680 & 148.73982 &  39.74292 & 0.198 & 10.46 &  9.21 &  7.73 & 19.64$\pm$0.15 & 19.27$\pm$0.08 &  2.61 & 11.08 &  1.26 & \nodata \\  
588009367476306065 & 148.80473 &  57.53416 & 0.141 & 10.40 &  2.46 &  8.82 & 18.99$\pm$0.12 & 19.22$\pm$0.08 &  0.19 &  9.62 &  0.82 &  8.60 \\    
588016526633861363 & 149.22546 &  38.67062 & 0.205 & 10.40 &  8.84 &  7.71 & 19.90$\pm$0.18 & 19.85$\pm$0.10 &  1.58 & 10.85 &  0.55 & \nodata \\  
588009367476568126 & 149.58009 &  57.82781 & 0.201 & 10.42 & 12.82 &  7.41 & 19.78$\pm$0.17 & 19.52$\pm$0.10 &  2.84 & 11.20 &  0.86 & \nodata \\  
587735239571603546 & 149.87546 &  38.15611 & 0.211 & 10.31 &  9.22 &  7.58 & 20.17$\pm$0.20 & 19.72$\pm$0.09 &  3.47 & 10.87 &  1.04 &  9.13 \\    
588009365865758909 & 150.09341 &  56.59479 & 0.228 & 10.31 & 10.50 &  7.47 & 20.37$\pm$0.22 & 20.08$\pm$0.13 &  2.20 & 11.03 &  1.08 & \nodata \\  
587735239571669282 & 150.10429 &  38.41888 & 0.252 & 10.41 & 12.81 &  7.40 & 20.36$\pm$0.22 & 20.37$\pm$0.13 &  3.54 & 11.13 &  1.33 &  9.08 \\    
587729387144806501 & 150.50368 &  55.12448 & 0.247 & 10.45 &  5.26 &  8.21 & 20.21$\pm$0.20 & 19.58$\pm$0.09 &  2.42 & 11.09 &  1.70 & \nodata \\  
588297864722382971 & 152.10156 &  42.73342 & 0.181 & 10.33 &  4.58 &  8.21 & 19.76$\pm$0.18 & 19.72$\pm$0.10 &  2.55 & 10.78 &  0.73 &  9.06 \\    
587732482202861656 & 152.16473 &  44.11267 & 0.151 & 10.56 & 11.73 &  7.62 & 18.76$\pm$0.10 & 18.55$\pm$0.05 &  0.59 & 11.05 &  0.64 & \nodata \\  
587732135914045580 & 152.51022 &  48.11020 & 0.255 & 10.44 & 15.28 &  7.27 & 20.33$\pm$0.22 & 20.13$\pm$0.16 &  2.05 & 11.40 &  0.65 & \nodata \\  
587732135377109127 & 152.62944 &  47.70260 & 0.190 & 10.46 &  7.01 &  7.97 & 19.56$\pm$0.15 & 19.40$\pm$0.08 &  2.44 & 10.78 &  1.22 &  9.07 \\    
587732136451047490 & 152.72443 &  48.49102 & 0.158 & 10.34 &  3.32 &  8.50 & 19.42$\pm$0.14 & 19.74$\pm$0.10 &  2.18 & 10.78 &  0.31 & \nodata \\  
587725816944459876 & 153.04662 &  63.41769 & 0.246 & 10.33 &  1.23 &  9.35 & 20.50$\pm$0.23 & 20.08$\pm$0.11 &  0.59 &  9.84 &  1.02 &  8.76 \\    
587732483814129906 & 153.54796 &  45.79028 & 0.161 & 10.38 & 11.82 &  7.44 & 19.37$\pm$0.14 & 18.97$\pm$0.06 &  1.80 & 11.02 &  0.81 & \nodata \\  
587729388219662459 & 153.73042 &  57.11367 & 0.255 & 10.35 & 10.92 &  7.48 & 20.53$\pm$0.24 & 19.93$\pm$0.10 &  4.46 & 11.21 &  0.93 &  9.15 \\    
587732152571986075 & 153.91460 &  45.34412 & 0.211 & 10.34 & 12.79 &  7.33 & 20.10$\pm$0.19 & 19.70$\pm$0.09 &  8.48 & 11.48 &  0.66 & \nodata \\  
588009368551817353 & 154.89285 &  60.22631 & 0.156 & 10.31 & 14.23 &  7.21 & 19.44$\pm$0.14 & 19.53$\pm$0.10 &  1.42 & 11.23 &  0.43 & \nodata \\  
587731500796739661 & 155.78958 &  54.49640 & 0.214 & 10.37 &  9.54 &  7.61 & 20.07$\pm$0.20 & 19.74$\pm$0.09 &  0.00 & 11.22 &  0.55 & \nodata \\  
587731869090709644 & 155.85841 &  51.10234 & 0.138 & 10.40 &  5.64 &  8.10 & 18.95$\pm$0.12 & 18.66$\pm$0.06 &  1.00 & 10.08 &  0.85 &  8.82 \\    
587732048949739624 & 156.04276 &  53.32828 & 0.185 & 10.40 & 10.56 &  7.55 & 19.65$\pm$0.16 & 19.28$\pm$0.08 &  0.69 & 11.03 &  0.82 &  9.05 \\    
587732048412868688 & 156.30435 &  52.87869 & 0.177 & 10.31 & 12.45 &  7.32 & 19.76$\pm$0.17 & 19.53$\pm$0.10 &  1.76 & 10.96 &  1.09 & \nodata \\  
587732582053117989 & 156.54250 &  56.66665 & 0.197 & 10.32 &  7.57 &  7.76 & 20.00$\pm$0.18 & \nodata &  4.31 & 11.24 &  0.62 & \nodata \\         
587732135378354188 & 156.55821 &  48.74970 & 0.160 & 10.47 &  2.17 &  9.00 & 19.13$\pm$0.12 & 18.91$\pm$0.06 &  0.25 &  9.91 &  1.10 &  8.33 \\    
588013383810220184 & 157.34006 &  49.69535 & 0.194 & 10.35 & 11.42 &  7.44 & 19.88$\pm$0.17 & 19.95$\pm$0.10 &  2.49 & 10.81 &  0.60 & \nodata \\  
587731870701977714 & 157.43727 &  52.86215 & 0.227 & 10.36 &  6.85 &  7.89 & 20.24$\pm$0.21 & 20.06$\pm$0.13 &  3.37 & 11.05 &  1.94 &  9.06 \\    
588009370685276243 & 157.76656 &  58.93386 & 0.173 & 10.44 &  2.46 &  8.86 & 19.37$\pm$0.17 & 18.72$\pm$0.07 & \nodata & 10.13 &  1.42 &  8.70 \\  
588007005245472903 & 157.82327 &  62.65335 & 0.198 & 10.39 &  4.69 &  8.25 & 19.83$\pm$0.17 & 20.10$\pm$0.13 &  1.40 & 10.90 &  0.70 & \nodata \\  
588007005782540350 & 158.33391 &  63.18797 & 0.145 & 10.62 &  5.32 &  8.37 & 18.50$\pm$0.09 & 18.18$\pm$0.05 &  2.25 & 10.99 &  2.10 &  9.06 \\    
587725551744909441 & 159.76553 &  65.55137 & 0.117 & 10.43 &  7.33 &  7.90 & 18.47$\pm$0.09 & 18.17$\pm$0.05 &  2.70 & 10.82 &  1.07 &  9.12 \\    
587732049487790084 & 159.83974 &  54.78919 & 0.186 & 10.57 &  2.92 &  8.84 & 19.21$\pm$0.13 & 18.62$\pm$0.05 &  0.09 &  9.54 &  1.13 &  8.62 \\    
587729386073489559 & 160.15704 &  56.81467 & 0.148 & 10.32 &  3.03 &  8.56 & 19.29$\pm$0.16 & 19.22$\pm$0.09 &  0.15 & 10.64 &  1.67 &  8.85 \\    
587725817483034723 & 160.69228 &  65.39616 & 0.189 & 10.34 &  9.44 &  7.59 & 19.84$\pm$0.05 & 19.18$\pm$0.02 &  1.59 & 11.04 &  1.45 &  9.13 \\    
587725551745564802 & 162.93958 &  66.10591 & 0.170 & 10.31 &  0.86 &  9.64 & 19.65$\pm$0.04 & 19.30$\pm$0.02 &  1.08 &  9.57 &  1.15 &  8.83 \\    
587725819094696047 & 165.08723 &  67.48546 & 0.191 & 10.33 &  5.02 &  8.13 & 19.90$\pm$0.23 & 19.42$\pm$0.10 &  1.50 & 11.19 &  1.37 &  9.14 \\    
587729386075455636 & 168.17336 &  58.13764 & 0.154 & 10.36 & 11.11 &  7.47 & 19.29$\pm$0.13 & 18.80$\pm$0.06 &  1.82 & 11.11 &  1.03 & \nodata \\  
587729153595539594 & 169.57059 &  63.26297 & 0.238 & 10.38 &  2.24 &  8.88 & 20.30$\pm$0.22 & 20.42$\pm$0.13 &  1.27 & 10.23 &  1.34 &  8.91 \\    
588011099410071613 & 169.80771 &  62.91817 & 0.286 & 10.32 &  4.04 &  8.31 & 20.90$\pm$0.28 & 20.27$\pm$0.12 &  0.75 & 10.95 &  0.67 & \nodata \\  
587729155743875234 & 173.26578 &  65.22815 & 0.241 & 10.72 &  0.38 & 10.77 & 19.48$\pm$0.04 & 19.62$\pm$0.02 &  0.01 &  8.94 &  0.51 &  8.19 \\    
588009372836757702 & 174.94957 &  63.15313 & 0.245 & 10.41 &  1.98 &  9.02 & 20.29$\pm$0.23 & 20.47$\pm$0.14 &  1.59 & 10.40 &  1.11 & \nodata \\  
587725816412766387 & 180.20340 &  66.43546 & 0.120 & 10.35 &  8.36 &  7.71 & 18.73$\pm$0.10 & 18.61$\pm$0.06 &  1.86 & 11.05 &  0.71 &  9.10 \\    
587728677395234987 & 189.59756 &  64.20047 & 0.212 & 10.38 & 10.17 &  7.57 & 20.00$\pm$0.21 & 19.87$\pm$0.11 &  0.25 & 11.12 &  0.97 & \nodata \\  
587728677932105821 & 189.88033 &  64.68499 & 0.133 & 10.50 &  1.48 &  9.36 & 18.60$\pm$0.11 & \nodata &  0.52 &  9.83 &  0.94 &  8.62 \\           
587725552827105291 & 208.48289 &  66.80015 & 0.198 & 10.66 &  1.89 &  9.31 & 19.16$\pm$0.13 & 18.48$\pm$0.05 &  0.84 &  9.99 &  1.71 &  8.72 \\    
588848899930194109 & 219.96780 &  -0.07335 & 0.203 & 10.34 & 10.99 &  7.46 & 20.01$\pm$0.08 & 19.77$\pm$0.04 &  2.37 & 10.91 &  0.69 & \nodata \\  
587725575888634103 & 260.11795 &  59.22827 & 0.221 & 10.46 & 12.54 &  7.47 & 19.92$\pm$0.06 & \nodata &  2.39 & 11.40 &  0.17 & \nodata \\         
587725575888634081 & 260.18839 &  59.27158 & 0.154 & 10.31 &  7.50 &  7.76 & 19.42$\pm$0.04 & 19.00$\pm$0.02 &  2.34 & 10.66 &  0.82 &  9.02 \\    
587725576425505123 & 261.11240 &  59.28925 & 0.255 & 10.50 &  9.42 &  7.75 & 20.17$\pm$0.06 & 19.83$\pm$0.04 &  0.84 & 11.31 &  0.96 & \nodata \\  
587730847960924833 & 312.50006 &   0.52355 & 0.164 & 10.37 &  0.60 & 10.01 & 19.42$\pm$0.23 & 19.67$\pm$0.18 &  1.47 & 10.65 &  0.93 & \nodata \\  
587730846887707478 & 313.79340 &  -0.31749 & 0.198 & 10.41 &  2.52 &  8.81 & 19.78$\pm$0.24 & 18.74$\pm$0.09 &  1.09 & 10.25 &  1.00 &  8.74 \\    
587730847962104313 & 315.19748 &   0.55589 & 0.254 & 10.59 &  6.32 &  8.19 & 19.93$\pm$0.24 & 19.85$\pm$0.14 &  1.26 & 11.25 &  1.33 & \nodata \\  
587730848499237044 & 315.84958 &   0.99881 & 0.177 & 10.32 &  7.51 &  7.77 & 19.72$\pm$0.31 & 19.53$\pm$0.17 &  4.22 & 11.15 &  1.51 & \nodata \\  
587727212271567298 & 317.70895 &  -8.01180 & 0.211 & 10.39 & 21.74 &  6.92 & 19.98$\pm$0.08 & 19.76$\pm$0.04 &  6.37 & 11.15 &  0.63 & \nodata \\  
587730847428510324 & 322.76953 &   0.09779 & 0.178 & 10.33 &  7.76 &  7.75 & 19.72$\pm$0.18 & 19.65$\pm$0.12 &  2.25 & 10.56 &  0.93 &  9.06 \\    
587730846891704703 & 322.89319 &  -0.30288 & 0.213 & 10.66 & 10.84 &  7.79 & 19.33$\pm$0.14 & 19.84$\pm$0.16 &  2.37 & 11.44 &  0.63 & \nodata \\  
587727212273992007 & 323.27859 &  -8.59399 & 0.217 & 10.51 &  9.72 &  7.74 & 19.76$\pm$0.19 & 20.17$\pm$0.13 &  3.08 & 11.16 &  0.70 & \nodata \\  
587731186725814690 & 325.83533 &   0.25139 & 0.169 & 10.37 & 13.48 &  7.31 & 19.50$\pm$0.19 & 19.16$\pm$0.13 &  0.00 & 11.37 &  0.60 & \nodata \\  
587731187799687683 & 326.06918 &   1.09551 & 0.243 & 10.48 & 11.41 &  7.57 & 20.11$\pm$0.23 & 19.99$\pm$0.17 &  0.85 & 11.06 &  1.12 &  9.27 \\    
587731187799753015 & 326.25107 &   1.19933 & 0.204 & 10.39 &  0.77 &  9.82 & 19.91$\pm$0.22 & 19.24$\pm$0.11 &  1.32 & 10.12 &  1.47 &  8.86 \\    
587726877808591211 & 333.18124 &  -9.00390 & 0.247 & 10.40 & 12.67 &  7.40 & 20.35$\pm$0.25 & 19.98$\pm$0.12 &  0.71 & 11.08 &  0.82 & \nodata \\  
587726879958302970 & 338.32794 &  -7.73724 & 0.210 & 10.39 & 11.18 &  7.49 & 19.97$\pm$0.23 & 19.38$\pm$0.15 &  1.09 & 11.02 &  0.90 & \nodata \\  
587726877811605723 & 340.03107 &  -9.63157 & 0.154 & 10.42 & 11.17 &  7.53 & 19.14$\pm$0.20 & 19.46$\pm$0.14 &  2.19 & 10.97 &  0.54 & \nodata \\  
587726877274800314 & 340.15994 & -10.09933 & 0.084 & 10.37 &  8.62 &  7.70 & 17.85$\pm$0.10 & 17.72$\pm$0.06 &  3.27 & 10.55 & -0.02 & \nodata \\  
587730815752667321 & 340.18524 & -10.27453 & 0.257 & 10.51 &  6.55 &  8.08 & 20.18$\pm$0.29 & 19.98$\pm$0.21 &  3.29 & 10.76 &  1.37 & \nodata \\  
587730817365245979 & 344.72656 &  -9.34214 & 0.116 & 10.50 &  3.40 &  8.64 & 18.30$\pm$0.09 & 17.84$\pm$0.04 &  0.67 &  9.97 &  1.15 &  8.65 \\    
587731187808731205 & 346.76562 &   1.21978 & 0.126 & 10.43 &  0.71 &  9.93 & 18.66$\pm$0.02 & 18.27$\pm$0.01 &  0.65 & 10.15 &  0.92 &  8.82 \\    
587731187809059012 & 347.51736 &   1.17908 & 0.207 & 10.41 &  6.18 &  8.03 & 19.88$\pm$0.20 & 19.75$\pm$0.13 &  2.19 & 11.10 &  1.06 &  9.16 \\    
588015508191051903 & 349.55417 &  -0.69060 & 0.252 & 10.85 &  2.40 &  9.29 & 19.27$\pm$0.04 & 18.82$\pm$0.02 &  0.76 &  9.92 &  1.56 &  8.68 \\    
587731186736365813 & 349.90442 &   0.22842 & 0.185 & 10.74 & 20.00 &  7.34 & 18.79$\pm$0.03 & 19.63$\pm$0.03 & \nodata & 10.85 &  0.50 & \nodata \\
587731187273892048 & 351.41345 &   0.75200 & 0.277 & 10.52 &  1.06 &  9.67 & 20.32$\pm$0.06 & 20.20$\pm$0.03 &  0.73 &  9.31 &  0.78 &  8.61 \\    
587731187811549347 & 353.21692 &   1.23279 & 0.187 & 10.33 &  6.85 &  7.86 & 19.84$\pm$0.24 & 20.03$\pm$0.18 &  1.48 & 10.33 &  0.64 &  9.06 \\    
587731186737938564 & 353.46399 &   0.28075 & 0.198 & 10.50 &  7.96 &  7.90 & 19.56$\pm$0.19 & 19.27$\pm$0.12 &  1.17 & 10.56 &  1.61 &  8.96 \\    
587731186738266349 & 354.30264 &   0.27763 & 0.212 & 10.51 &  8.15 &  7.89 & 19.69$\pm$0.19 & 19.60$\pm$0.10 &  2.36 & 10.37 &  0.94 &  8.99 \\    
588015509266890925 & 354.41162 &   0.05134 & 0.186 & 10.31 &  8.35 &  7.67 & 19.87$\pm$0.20 & 20.39$\pm$0.15 &  3.09 & 10.59 &  0.45 & \nodata \\  
588015510342402184 & 358.44873 &   0.90061 & 0.223 & 10.48 &  2.17 &  9.01 & 19.89$\pm$0.19 & 20.03$\pm$0.14 &  0.27 &  9.55 &  0.53 &  8.22 \\    
588015510342468082 & 358.57800 &   0.94167 & 0.232 & 10.42 & 11.38 &  7.51 & 20.13$\pm$0.22 & 20.23$\pm$0.16 &  1.07 & 10.71 &  1.04 &  9.07 \\    
587727225152733191 & 358.70636 & -10.97845 & 0.121 & 10.35 &  1.89 &  9.00 & 18.76$\pm$0.12 & 18.38$\pm$0.06 &  0.74 & 10.08 &  0.94 &  8.62 \\    
588015509269053589 & 359.30862 &   0.14575 & 0.262 & 10.38 &  7.69 &  7.81 & 20.53$\pm$0.26 & 19.56$\pm$0.12 &  2.32 & 11.01 &  1.59 &  8.89 \\    

\enddata
\tablecomments{UVLG sample. Col$.$ (1): SDSS Object ID number. Col$.$ (2): Right Ascension (2000). Col$.$ (3): Declination (2000).  Col$.$ (4): Redshift from SDSS pipeline.  Col$.$ (5): Log of FUV luminosity (corrected for Galactic extinction).   Col$.$ (6): Seeing-corrected half light radius in $u$-band.  Col$.$ (7): Log of FUV surface brightness (corrected for Galactic extinction).   Col$.$ (8): FUV magnitude (corrected for Galactic extinction). Col$.$ (9): NUV magnitude (corrected for Galactic extinction). Col$.$ (10): FUV attenuation using the Balmer decrement measured on the SDSS spectra, and the Calzetti (2001) starburst attenuation law. Col$.$ (11): Log of stellar mass determined via SED fitting.  Col$.$ (12): Log of extinction-corrected star formation rate determined via SED fitting.   Col$.$ (13): Metallicity using the Tremonti et al (2004) method.}
\end{deluxetable}

\begin{deluxetable}{lcccc}
\tabletypesize{\scriptsize}
\tablecaption{Comparison of Galaxy Properties \label{proptab}}
\tablewidth{0pt}
\tablehead{
\colhead{Parameter} & \colhead{Large UVLGs} & \colhead{Compact UVLGs} & \colhead{Supercompact UVLGs} & \colhead{LBGs\tablenotemark{a}} \\
\colhead{} & \colhead{} & \colhead{($I_{1530}>10^8$~$L_{\odot}$~kpc$^{-2}$)} & \colhead{($I_{1530}>10^9$~$L_{\odot}$~kpc$^{-2}$)} & \colhead{}}
\startdata
Number                                  & 110              & 105\tablenotemark{b} & 35\tablenotemark{b} & \nodata \\
log $L_{1530}$ ($L_{\odot}$)            & $10.3$ to $11.2$ & $10.3$ to $11.0$ & $10.3$ to $10.9$    & $10.3$ to $11.3$ \\
log $I_{1530}$ ($L_{\odot}$ kpc$^{-2}$) & $6.0$ to $8.0$   & $8.0$ to $10.3$  & $9.0$ to $10.3$     & $9$ to $10$      \\
log $R_{50,u}$ (kpc)                    & $0.9$ to $1.6$   & $-0.5$ to $0.8$  & $-0.5$ to $0.4$     & $0.0$ to $0.5$   \\
log $M_*$ ($M_{\odot}$)                 & $10.3$ to $11.7$ & $9.2$ to $11.0$  & $9.0$ to $10.7$     & $9.5$ to $11.0$  \\
$A_{1530}$                              & $0$ to $5$       & $0$ to $2.5$       & $0$ to $2$          & $0$ to $3$     \\
log SFR ($M_{\odot}$ yr$^{-1}$)         & $0$ to $1.5$     & $0.2$ to $2$     & $0.5$ to $2$        & $0.5$ to $2.5$   \\
log SFR/$M_*$ (yr$^{-1}$)               & $-11$ to $-9.5$  & $-10.5$ to $-8$  & $-9.3$ to $-8$      & $-9$ to $-8$     \\
FUV$ -r$                                & $1.0$ to $3.5$   & $0.2$ to $2.8$   & $0.2$ to $1.7$      & $0.2$ to $2.2$   \\
12+log(O/H)\tablenotemark{c}            & $8.6$ to $9.3$   & $8.5$ to $9.2$   & $8.2$ to $8.9$      & $7.7$ to $8.8$   \\ 
12+log(O/H)\tablenotemark{d}            & $8.4$ to $8.9$   & $8.2$ to $8.8$   & $8.1$ to $8.8$      & $8.2$ to $8.6$   \\ 
\enddata
\tablenotetext{a}{Properties for LBGs are taken from Shapley et al. (2001), Erb et al. (2006a), Papovich et al. (2001), Giavalisco et al (2002), and Ferguson et al. (2004).}
\tablenotetext{b}{The supercompact UVLGs are a subset of the compact UVLGs, {\it i.e.,} the compact UVLG sample includes the 35 supercompact UVLGs.}
\tablenotetext{c}{Metallicity determined using the Tremonti et al (2004) technique.}
\tablenotetext{d}{Metallicity determined using the $N2$ technique (see Pettini \& Pagel 2004; Erb et al 2006a.)\\}
\end{deluxetable}


\end{document}